\documentclass[superscriptaddress,aps,prd,showpacs]{revtex4}
\usepackage{amsfonts}
\usepackage{graphicx}
\usepackage{amsmath}
\usepackage{color}

\begin{document}

\title{Electro- and magneto-statics of topological insulators as modeled by planar, spherical and cylindrical $\theta$ boundaries: Green's function approach}

\author{A. Mart\'{i}n-Ruiz}
\email{alberto.martin@nucleares.unam.mx}
\affiliation{Instituto de Ciencias Nucleares, Universidad Nacional Aut\'{o}noma de M\'{e}xico, 04510 M\'{e}xico, Distrito Federal, M\'{e}xico}

\author{M. Cambiaso}
\affiliation{Universidad Andres Bello, Departamento de Ciencias Fisicas, Facultad de
Ciencias Exactas, Avenida Republica 220, Santiago, Chile}

\author{L. F. Urrutia}
\affiliation{Instituto de Ciencias Nucleares, Universidad Nacional Aut\'{o}noma de M\'{e}%
xico, 04510 M\'{e}xico, Distrito Federal, M\'{e}xico}

\date{\today }

\begin{abstract}
The Green's function (GF) method is used to analyze the boundary 
effects produced by a Chern-Simons (CS) extension to 
electrodynamics. We consider the electromagnetic field coupled
to a $\theta$ term that is piecewise constant in different regions
of space, separated by a common interface $\Sigma$, the 
$\theta$ boundary, model which we will refer to as 
$\theta$ electrodynamics ($\theta$ ED). 
This model  provides a correct low energy effective
action for describing topological insulators (TI).
Features arising due  to the presence of the
boundary, such as  magnetoelectric effects, are already 
known in CS extended electrodynamics and  solutions 
for some  experimental setups have been found  with 
specific configuration of sources. 
In this work we
construct the static GF  in $\theta$ ED for different 
geometrical configurations of the $\theta$ boundary, namely: planar, spherical and cylindrical $\theta$-interfaces. Also we
adapt
the standard Green's theorem to include the effects of the 
$\theta$ boundary.  
These are the most important results of our work, since they allow to obtain  the corresponding static electric and magnetic fields for arbitrary sources and arbitrary boundary conditions in the given geometries. Also, the method provides a well defined  starting point for either analytical or  numerical approximations in the cases where the exact analytical calculations are not possible.
Explicit solutions for simple cases in each of 
the aforementioned geometries for $\theta$ boundaries are provided.
On the one hand, the adapted Green's theorem is illustrated 
by studying
the problem of a point-like electric charge interacting
with a planar TI  with prescribed  boundary conditions. On the other hand we calculate the electric and magnetic static fields 
produced by the following sources: (i) a point-like
electric charge near a spherical $\theta$ boundary, (ii) an infinitely straight
current-carrying wire near a cylindrical $\theta$ boundary and (iii) an infinitely straight uniformly-charged 
wire near a cylindrical $\theta$ boundary. 
Our generalization, when particularized to specific cases, is successfully compared with  previously reported results, most of which have been obtained by using the methods of images.
\end{abstract}

\maketitle

\section{Introduction}

\label{introduction}

It is not seldom that seemingly abstract mathematical models find widespread
application in several fields of theoretical as well as applied physics. 
A paramount example of this is Einstein's theory of general relativity,
because in order to achieve the accuracy that render GPS useful,
relativistic effects must be taken into account. Filling the gap from theory
to applied engineering is only a matter of time. Fortunately, the time span
from conception to application gets shorter and shorter. So is the case with
the study of Chern-Simons (CS) forms \cite{Chern:1974ft} to its applications
in topological insulators, spintronics and topological quantum computer
science \cite{tech-apps,TA-2}.

The relevance of CS forms also is apparent in theoretical physics. In
quantum field theories they play a prominent role in regards to anomalies. 
The existence of anomalies can jeopardize the consistence of the theory as
they would be indicative of gauge symmetry violation, hence the need for
anomaly cancellation mechanisms. Furthermore, anomalies are needed to
account for certain experimental observables \textit{e.g.}, the proper
neutral pion decay rate into two photons as predicted by the
non-conservation of the axial-vector current generated by the ABJ anomaly %
\cite{Bell:1969ts, Adler:1969gk}.
The latter can be expressed as the one-loop
contribution to the divergence
of a pseudo-vector or axial current $\partial^\mu j_{\mu A} \propto {}^{\ast}FF$. In general relativity a similar argument holds. In $3+1$
spacetime dimensions general relativity can be thought of as a non-abelian
gauge theory for the $SO(3,1)$ gauge group of local Lorentz transformations,
for which case a gravitational anomaly exists $D^\mu J_{\mu \ell} \propto 
\mathcal{P}$, where $J_{\mu \ell}$ is the lepton-number current and $%
\mathcal{P} = {}^{\ast}RR$ is the Pontryagin class in $3+1$ dimensions. Here
one can cancel the anomaly by adding the appropriate counter term to the
Einstein-Hilbert action, which turns out to be a CS extension of general
relativity. %
Further, this CS extended general relativity when applied to the
cosmological context was posed as a possible explanation for
matter-antimatter (baryon) asymmetry in the Universe \cite{Alexander:2004us}%
. Actually it is a lepton number asymmetry, however particular weak
interaction processes in the SM mediated by $SU(2)$ instantons (sphalerons)
can transmute leptons into baryons under certain conditions that would have
been met during the out-of-equilibrium inflationary epoch of the Universe %
\cite{Kuzmin:1985mm,Fukugita:1986hr}.

In a very simple guise, CS terms were used by Peccei and Quinn by the
introduction of the axion in order to solve the strong CP-problem of QCD %
\cite{Peccei:1977hh}. Soon after 't Hooft realized that the unobserved $U(1)$
symmetry of QCD can be understood due to the dynamics of instantons \cite%
{'tHooft:1986nc}. Wilczek explicitly predicted applications of
axion-electrodynamics to the field of material science \cite{Wilczek:1987mv}.  
Further studies involve its uses in: topological quantum field theory \cite
{Witten:1988ze}, topological string theory \cite{Marino:2004uf} and as a
quantum gravity candidate \cite{Bonezzi:2014nua}.

In this work we will be concerned with a simple case of CS theories, akin to 
axion-electrodynamics, introduced by Wilczek as mentioned above in the context
of particle physics. We will refer to this particular model as 
$\theta$-electrodynamics or simply $\theta$ ED and it amounts to extending Maxwell
electromagnetism by a gauge invariant 
term of the form 
\begin{equation}  \label{EdotB}
\Delta \mathcal{L}_\theta = \theta (\alpha / 4 \pi^2) \mathbf{E} \cdot 
\mathbf{B}.
\end{equation}
Here though, $\theta$ is no longer a dynamical field but rather we take it
as a constant, a genuine Lorentz scalar and thus Eq.~(\ref{EdotB}) is a
pseudo-scalar. Written in a manifestly covariant way 
\begin{equation}  \label{FF*}
\Delta \mathcal{L}_\theta =-\frac{\theta}{4} (\alpha / 4 \pi^2) F_{\mu \nu}
\tilde F^{\mu \nu} ,
\end{equation}
the identification with the Pontryagin invariant associated with the $U(1)$
gauge connection $A = A_\mu dx^\mu$ is immediate. In the latter we
introduced the Hodge dual field strength electromagnetic tensor $\tilde{F}%
^{\mu \nu }=\frac{1}{2}\epsilon ^{\mu \nu \alpha \beta }F_{\alpha \beta }$
and $\epsilon ^{\mu \nu \alpha \beta }$ is the Levi-Civita symbol.

The idea of $\theta$ ED has been studied in several situations and also
extrapolated to study other systems. In other contexts, $\theta$ ED 
under consideration here has been studied as a 3+1 particular kind of
Maxwell-Chern-Simons electrodynamics (MCS), where it has received
considerable attention \cite{Carroll:1989vb, Milton:1990yj, Milton:1992rf,
Devecchi:1994ha, Anacleto:2001rp, Blasi:2010gw, Ozonder:2010zy,
Tallarita:2010vu, Mukherjee:2011da, Andrade:2011sx, ConchaSanchez:2013cp,
Chang:2014jna, Balachandran:1993tm, Asorey:2015sra, Asorey:2013wvh, Markov:2006, Wegner:2015}. And also it has been studied as a restricted subset of the
Standard Model Extension (SME) \cite{Colladay:1996iz,Colladay:1998fq}, where
several results have been achieved too \cite{Belich:2003xa, Lehnert:2005rh,
Alfaro:2006dd, Hariton:2006zj, Kaufhold:2007qd, Kharlanov:2009pv,
Alfaro:2009mr}.

Further, the topological nature of the $\theta$ term of Eq. (\ref{FF*}) 
can be seen by the fact that this CS
extension is a total derivative. Therefore it produces no contribution to
the field equations of motion when usual boundary conditions are met, the
contribution is a boundary term that vanishes whenever one imposes the
vanishing of the fields at the boundary (or at infinity in the case the
theory is defined over the whole space). Should $\theta$ cease to be a
constant in the manifold where the theory is defined, the CS term
fails to be a topological invariant and therefore the corresponding
modifications to the field equations must be taken into consideration.

In this paper we will be concerned with the simplest non-trivial case in
this context, namely we will study the modifications to Maxwell's theory
defined on a manifold in which, either: (a) there are two domains defined by
their different constant values of $\theta$ \textit{i.e.}, $\theta$ is space dependent $\theta(x)$
or (b) there is a non-vanishing $\theta$
value and the manifold has a boundary where the fields or their derivatives
are not vanishing, with either Dirichlet or Neumann boundary
conditions. A constant $\theta$  can be thought of as an
effective parameter characterizing properties of a novel electromagnetic
vacuum, possibly arising from a more fundamental theory, where discontinuities in
the value of $\theta$ has interesting properties. 
This approach 
 has been taken in
the context of classical $\theta$ ED \cite{science, Hehl-Obukov, Huerta:2012ks, Huerta:2014ula}
and in the context of quantum vacuum \cite{Canfora:2011fd}. 
A similar avenue has been taken in the context of Janus 
field theories \cite{CFKS,DEG,Chen,Witten2,Kim1,Kim2}. 
These were motivated from the gravitational sector of the AdS/CFT correspondence,  by an exact and  non-singular  solution for the dilatonic field in  type IIB supergravity
\cite{Bak}. The similarity owing to  the fact that the coupling constant of the 
ensuing 4-dimensional $\mathcal{N} = 4$ super Yang-Mills  theory living in the boundary exhibits a space-time
dependent character. For further 
insight on the similarities and differences between Janus field theories and
$\theta$ ED as studied in this work, see \cite{Martin-Ruiz:2015skg}
 and references therein.
On the other hand, as applied to
material media, $\theta$ can be regarded as an effective macroscopic 
parameter to describe new 
degrees of freedom of quantum matter. This approach has been thoroughly 
used in the context of topological
insulators (TI) as will be explained below.

First, recall that in general CS forms are amenable for capturing topological features of the
physical system that they describe.
Formally this can be seen from the action principle. Given a symmetry group
and an odd-dimensional differentiable manifold where fields and functions
are defined, a gauge connection one-form can be defined whose associated
curvature two-form can be used to build $2k$-forms that: (i) are gauge
invariant under the symmetry group, (ii) are closed and therefore
expressible in terms of a $(2k-1)$-form and (iii) its integral is a
topological invariant. This last point is crucial revealing the importance
of boundaries. Its many uses in gravitation and the former description are
clearly reviewed in \cite{Zanelli:2012px}.

The latter description is tailor-made for the understanding what came to be
known as topological phases. The discovery of the Quantum Hall (QH) state
made manifest the existence of new states of matter that do not fall into
Landau-Ginzburg's effective field theory paradigm. 
In it the quantum
mechanical states of matter that determine the different phases are
characterized by the spontaneous breaking of a global symmetry of the
quantum mechanical system. von Klitzing's discovery of the astonishing
precision with which the Hall conductance of a sample is
quantized \cite%
{von Klitzing:1980kg}, despite the varying irregularities of the sample,
turned out to have a topological origin. The ensuing electric current along
the edge of a $2D$ electron gas at very low temperature, due to an external
magnetic field applied perpendicular to the sample, is in fact insensitive
to the sample's geometric details. The reason for this lies in the band
structure of the sample. For the QH state the system is insulating in the
bulk and conducting in the boundary. The Hamiltonian of the many-particle
system in the bulk exhibits an energy gap separating the ground state from
the excited states. On the contrary, for the edge states the band structure
is gapless. Furthermore one can define a smooth deformation in the
Hamiltonian's parameter space (the symmetry transformation referred to in
the previous paragraph, the transformation taking the coffee mug to a torus)
that does not close the bulk gap. To finally understand the connection with
topology, one can recall the Gauss-Bonnet formula and Berry's phase, the
generalization of the concept of curvature to quantum mechanical systems.
The former allows to express the genus $g$ of a surface $S$ in terms of an
integral over the local curvature of the surface. The latter is a measure of
the phase accumulated by a wave-function as it evolves under a slow and
closed variation in the parameter space of the Hamiltonian. Back in the QH
scenario, the Hall conductance can be expressed as an invariant integral
over the frequency momentum space, more precisely as an integral of the
Berry curvature over the Brillouin zone \cite{SQ Shen}. This quantity plays
the role of a topological order parameter uniquely determining the nature of
the quantum state inasmuch as the order parameter in Landau-Ginzburg
effective field theory determines the usual phases of quantum matter.

As mentioned previously, CS terms lends themselves for the description of
topological features of a given physical system. Concretely, in material
science systems, the low-energy limit of the electrodynamics of topological
insulators can be described by extending Maxwell electrodynamics precisely
by the $\theta$ term of Eq. (\ref{FF*}), originally formulated in 4+1
dimensions but appropriately adapted to lower dimensions by dimensional
reduction \cite{Qi:2008ew}. Thus, $\theta$ ED as a topological field theory
(TFT), serves as model for many theoretical \cite%
{Maciejko:2010yg,Maciejko:2010tx,Wang:2010uy} and experimental realizations
for studying detailed properties of topological states of quantum matter
\cite{tech-apps,Xu:2015cga,Qi:2008pi,Ge:2015fk}. 
See also the review papers \cite{Qi:2011zya,Hasan:2010xy} and references therein.

The general scope of this work is to introduce Green's function (GF)
methods in $\theta$ ED, which are well suited to deal with the calculation
of electric and magnetic fields arising from arbitrary sources, as well as
to solve problems with given Dirichlet or Neumann boundary conditions on
arbitrary surfaces. This approach is
more general than the 
method of images
which, to our knowledge, has been systematically employed in most of the previous 
works on the related literature.
On the other hand,  the GF method provides a precise starting point from where either
analytical or numerical approximations can be performed. In Ref. 
\cite{Martin-Ruiz:2015skg}
we have already presented the first steps in this direction by constructing
the GF for a $\theta$ boundary with planar geometry. The method was applied to the
calculation of some specific examples. The GF for a stack of layered time-reversal-symmetry-broken TI has been constructed in Ref. \cite{Crose} and applied to describe novel field patterns arising from the magnetoelectric effect due to  a dipole close to the surface of the TI.

The paper is organized as follows. In section \ref{model} we review the
basics of Chern-Simons electrodynamics defined on a four dimensional
spacetime characterized by a piecewise constant  value of  $\theta$ in different
regions of space separated by a common boundary $\Sigma$. 
To isolate the effects of the $\theta$-term on the GFs and of the stress-energy tensor,
we will take the media on either side of the $\Sigma$ boundary with no dielectric nor
magnetic properties, \textit{i.e.}, $\epsilon=1$ and $\mu=1$ across $\Sigma$.  Including 
electric and magnetic susceptibility properties proceeds accordingly.
We will restrict
our analysis to the case of electro- and magneto-statics. As in usual Maxwell
electrodynamics a robust static theory is necessary and important to
understand the dynamical case as well as the ensuing quantum
electrodynamics. In this scenario the field equations remain the standard
Maxwell equations in the bulk, but the discontinuity of $\theta$ modifies
the behavior of the fields at the interface $\Sigma$. The most striking
feature of this theory is that even in the static limit, electric and
magnetic fields are intertwined.
Aspects about a  
possibility to circumvent Earnshaw's theorem and 
 how to obtain the modified conservation laws are revised, in particular a detailed construction of the
stress-energy tensor is presented.

In section \ref{Green} we start by 
adapting Green's
theorem to  incorporate
the contributions of the $\theta$ boundary. Also we
classify the different boundary conditions that can be imposed there.
Restricting ourselves to the case where we do not impose boundary conditions at the $\theta$ interface (\textit{i.e.}, boundary
conditions only at infinity) we consider the simplest geometries for the 
$\Sigma$: planar, cylindrical and spherical. We present
a brief review of the planar case and construct the static GF matrix for the remaining 
 geometries, thus providing the general solution to the modified
field equations for the case of arbitrary configuration of sources. The
method can be extended to the case of other geometries but we focus on these
configurations given the fact that those are the ones that have attracted
the most part of the experimental efforts. This is an important part of our
paper. In it, not only do we aim to provide practical solutions for a given
boundary value problems with arbitrary external sources, but also to provide a means to improve the understanding of the theory.

In the appendices we present the detailed calculations of the Green's matrix elements for
the cases of cylindrical and  spherical geometries for the $\theta$ boundary. 
In Section \ref{applications} we present a simple
application of the adapted Green's theorem when Dirichlet boundary
conditions are imposed on a planar $\Sigma $. Also we discuss other
applications, where we make contact between the results obtained with our
method and others in the existing literature, a comparison that endorses our
approach. For instance, the problem of a point-like charge near a spherical $%
\theta$ boundary together with the cases of a current-carrying wire and a
uniformly charged
wire near a cylindrical $\theta$ boundary.
Finally, in section \ref{summary} we summarize our results and give further concluding remarks.
Throughout
the paper, Lorentz-Heaviside units are assumed ($\hbar = c = 1$), the metric
signature will be taken as $\left( + , - , - , - \right)$ and the convention 
$\epsilon^{0123} = +1$ is adopted.

\section{$\protect\theta$ Electrodynamics in a bounded region}

\label{model}

\subsection{Field equations and boundary conditions}

\label{FEandBC}

Let us consider the additional coupling of electrodynamics with the
electromagnetic Pontryagin invariant 
\begin{equation}
\mathcal{P}=F_{\mu \nu }\tilde{F}^{\mu \nu },
\end{equation}
via a scalar field $\theta $ defined by the action
\begin{equation}
\mathcal{S}=\int_{\mathcal{M}}d^{4}x\left[ -\frac{1}{16\pi }F_{\mu \nu
}F^{\mu \nu }-\frac{1}{4}\frac{\theta (x) }{2\pi }\frac{\alpha }{2\pi }F_{\mu
\nu }\tilde{F}^{\mu \nu }-j^{\mu }A_{\mu }\right] ,  \label{action}
\end{equation}
where $\alpha =e^{2}/\hbar c$ is the fine structure constant, $\tilde{F}^{\mu \nu }=\frac{1}{2}\epsilon ^{\mu \nu \alpha \beta }F_{\alpha \beta }$
is the dual of the field strength tensor and $j^{\mu }$ is a conserved
external current. 
The coupling constant for the $\theta$-term, $\alpha / 4 \pi ^{2}$, is
chosen in such a way that the Dirac quantization condition for the magnetic
charge be an integer multiple of $e / 2 \alpha$, as discussed in \cite%
{Wilczek:1987mv}. The topological nature of the Pontryagin density makes the equations of motions arising from the action (\ref{action}) invariant under the change $\theta (x) \rightarrow \theta (x) + C$, with $C$ any constant. Quantum mechanical arguments impose further conditions on $C$, which will be discussed in the following.

\begin{figure}[tbp]
\begin{center}
\includegraphics
{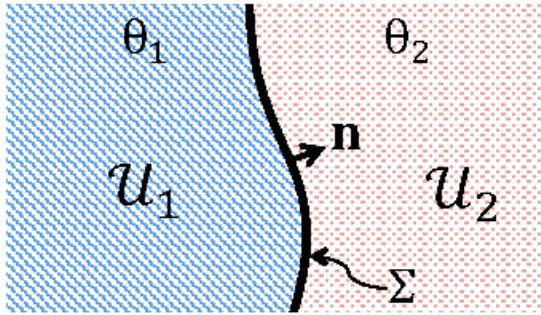}
\end{center}
\caption{{\protect\small Region over which the electromagnetic field theory is
defined.}}
\label{regions}
\end{figure}

The $(3+1)$-dimensional spacetime is $\mathcal{M}=\mathcal{U}\times \mathbb{R%
}$, where $\mathcal{U}$ is a three-dimensional manifold and $\mathbb{R}$
corresponds to the temporal axis. We make a partition of space in two
regions: $\mathcal{U}_{1}$ and $\mathcal{U}_{2}$ in such a way that
manifolds $\mathcal{U}_{1}$ and $\mathcal{U}_{2}$ intersect along a common
two-dimensional boundary $\Sigma $, to be called the $\theta $ boundary, so
that $\mathcal{U}=\mathcal{U}_{1}\cup \mathcal{U}_{2}$ and $\Sigma =\mathcal{%
U}_{1}\cap \mathcal{U}_{2}$, as shown in Fig.~ \ref{regions}. We also assume
that the field $\theta $ is piecewise constant in such way that it takes the
constant value $\theta =\theta _{1}$ in region $\mathcal{U}_{1}$ and the
constant value $\theta =$ $\theta _{2}$ \ in region $\mathcal{U}_{2}$. This
situation is expressed in the characteristic function 
\begin{equation}
\theta \left( \textbf{x} \right) =\left\{ 
\begin{array}{c}
\theta _{1}\;\;\;,\;\;\; \textbf{x} \in \mathcal{U}_{1} \\ 
\theta _{2}\;\;\;,\;\;\; \textbf{x} \in \mathcal{U}_{2}%
\end{array}%
\right. .  \label{theta}
\end{equation}%
In this scenario the $\theta$-term in the action fails to be a global
topological invariant because it is defined over a region with the boundary $%
\Sigma $. Varying the action gives rise to a set of Maxwell's equations with
an effective additional current with support at the boundary 
\begin{equation}
\partial _{\mu }F^{\mu \nu }=\tilde{\theta}\delta \left( \Sigma \right)
n_{\mu }\tilde{F}^{\mu \nu }+4\pi j^{\nu },  \label{FieldEqs}
\end{equation}%
where $n_{\mu } = (0,\textbf{n})$, $\textbf{n}$ is the outward unit normal to $\Sigma$ and 
\begin{equation}
\tilde{\theta}
= \frac{\alpha}{\pi} \left( \theta _{1}-\theta _{2}\right) . \label{Tilde}
\end{equation}
Current conservation can be verified directly by taking the divergence at
both sides of Eq. (\ref{FieldEqs}), 
\begin{equation}
\partial _{\nu }\partial _{\mu }F^{\mu \nu }=\tilde{\theta}\delta ^{\prime
}\left( \Sigma \right) n_{\mu }n_{\nu }\tilde{F}^{\mu \nu }+\tilde{\theta}%
\delta \left( \Sigma \right) n_{\mu }\partial _{\nu }\tilde{F}^{\mu \nu
}+4\pi \partial _{\nu }j^{\nu },  \label{Current}
\end{equation}
where the left hand side and the first two terms in the right hand side
vanish due to symmetry properties.

Together with the Bianchi identity $\partial _{\mu }\tilde{F}^{\mu \nu }=0$,
the set of Eqs. (\ref{FieldEqs}) for $\theta $ ED can be written in
coordinates adapted to the surface as 
\begin{eqnarray}
\nabla \cdot \mathbf{E} &=&\tilde{\theta}\delta \left( \Sigma \right) 
\mathbf{B}\cdot \mathbf{n}+4\pi \rho ,  \label{GaussE} \\
\nabla \times \mathbf{B}-\frac{\partial \mathbf{E}}{\partial t} &=&\tilde{%
\theta}\delta \left( \Sigma \right) \mathbf{E}\times \mathbf{n}+4\pi \mathbf{%
J},  \label{Ampere} \\
\nabla \cdot \mathbf{B} &=&0,  \label{GaussB} \\
\nabla \times \mathbf{E}+\frac{\partial \mathbf{B}}{\partial t} &=&0,
\label{Faraday}
\end{eqnarray}%
where $\mathbf{n}$ is the unit normal to $\Sigma $ shown in the Fig. \ref%
{regions}. In this work we consider the simplest geometries corresponding to
the cases where the surface $\Sigma $ is taken as: (i) the plane $z=a$, (ii) a sphere with center at the origin and radius $a$ and (iii) an infinitely straight
cylinder of radius $a$ with axis parallel to the $z $ direction.
In these cases the choice of adapted coordinates has an obvious meaning: (i)
Cartesian coordinates, (ii) spherical
coordinates and (iii) cylindrical coordinates, respectively.

As we see from Eqs. (\ref{GaussE}-\ref{Faraday}) the behavior of $\theta $ ED
in the bulk regions $\mathcal{U}_{1}$ and $\mathcal{U}_{2}$ is the same as
in standard electrodynamics. The $\theta $-term modifies Maxwell's equations
only at the surface $\Sigma $. Here $F^{i0}=E^{i}$, $F^{ij}=-\varepsilon
^{ijk}B^{k}$ and $\tilde{F}^{i0}=B^{i}$, $\tilde{F}^{ij}=\varepsilon
^{ijk}E^{k}$. Equations (\ref{GaussE}-\ref{Faraday}) also suggest that the
electromagnetic response of a system in the presence of a $\theta $-term can
be described in terms of Maxwell equations in matter 
\begin{equation}
\mathbf{\nabla }\cdot \mathbf{D}=4\pi \rho ,\;\;\;\mathbf{\nabla }\times 
\mathbf{H=}4\pi \mathbf{J},   \label{MEMATT}
\end{equation}%
with constitutive relations 
\begin{equation}
\mathbf{D}=\mathbf{E}+\frac{\alpha }{\pi }\theta \left( \textbf{x} \right) \mathbf{B}%
\;\;\;\;,\;\;\;\;\mathbf{H}=\mathbf{B}-\frac{\alpha }{\pi }\theta \left(
\textbf{x} \right) \mathbf{E},  \label{CONST_REL}
\end{equation}
where $\theta \left( \textbf{x} \right) $ is given in Eq. (\ref{theta}). It is clear
that if $\theta (\textbf{x})$ is globally constant in $\mathcal{M}$, there is no
contribution to Maxwell's equations from the $\theta$ term in the action,
even though $\theta (\textbf{x})$ still is present in the constitutive relations.
In fact, the additional contributions of a globally constant $\theta \left(
\textbf{x} \right) $\ to each of the equations (\ref{MEMATT}) cancel due to the
homogeneous equations (\ref{GaussB}) and (\ref{Faraday}).

So far we have considered $\theta$ as an external parameter that can take arbitrary values, albeit respecting the re-scaling symmetry $\theta \to \theta +C$ that we already mentioned. Naively this would allow to set $\theta$ to zero at the classical level. However, quantum mechanically, given that for properly quantized electric and magnetic fluxes $S _{\theta} / \hbar$ is an integer multiple of $\theta$, then the only allowed values of $C$ are $C = 2 \pi n$ for integer $n$, otherwise non-trivial contributions to the path integral would result. 
The above argument together with additional considerations  about the 
behavior of the system under  time-reversal (TR) can further constrain $\theta$. 
In fact,  in the context of TIs it was originally assumed that only  systems with broken 
TR symmetry were of relevance, however it was soon realized that systems preserving TR 
symmetry were as important, see Ref. \cite{Bernevig_book}. 
Given the pseudo-scalar nature under TR of the CS coupling:
$T( \mathbf{E} \cdot \mathbf{B} ) = - \mathbf{E} \cdot \mathbf{B}$, where 
$T$ is the time-reversal operator $T: t \to -t$,
it seems that breaking TR symmetry is inescapable. 
However, TR symmetry can be preserved  as long as $\theta$ takes the 
values $0$ or $\pi$ (mod $2 \pi$). For $\theta=0$ the system is trivially TR invariant 
whereas  
for $\theta=\pi$ it can be restored by 
suitably re-scaling $\theta$, \textit{i.e.,  }
$T( \pi \mathbf{E} \cdot \mathbf{B} ) = -  \pi \mathbf{E} \cdot \mathbf{B}$
can be made TR invariant by $\theta \to \theta' = \theta +2 \pi $.
For the sake of generality, in the sequel we will not restrict to 
systems with a  definite TR symmetry property,  therefore the only 
relevant constraint for $\theta$ that we will consider is that its observable effects
satisfy the $\theta \to \theta +2\pi n$ symmetry, \cite{Qi:2011zya}.
This property is made self-evident because all our observable results will depend only on the combination $\tilde{\theta}$ given in Eq. (\ref{Tilde}).

Assuming that the time derivatives of the fields are finite, in the vicinity
of the surface $\Sigma $ the field equations imply that the normal component
of $\mathbf{E}$, and the tangential components of $\mathbf{B}$, acquire
discontinuities additional to those produced by superficial free charges and
currents, while the normal component of $\mathbf{B}$, and the tangential
components of $\mathbf{E}$, are continuous. For vanishing external sources
on $\Sigma $ the boundary conditions read: 
\begin{eqnarray}
\Delta \mathbf{E}_{n}\big|_{\Sigma } &=&\tilde{\theta}\mathbf{B}_{n}\big|%
_{\Sigma },  \label{GaussE-BC} \\
 \Delta \mathbf{B}_{\parallel }\big|_{\Sigma } &=&-\tilde{\theta}\mathbf{E}%
_{\parallel }\big|_{\Sigma },  \label{Ampere-BC} \\
\Delta \mathbf{B}_{n}\big|_{\Sigma } &=&0,  \label{GaussB-BC} \\
\Delta \mathbf{E}_{\parallel }\big|_{\Sigma } &=&0.  \label{Faraday-BC}
\end{eqnarray}
The notation is $\mathbf{V}_{i}\big|^{z=a^+}_{z=a^-}={\mathbf V}_{i}(z)\big|^{z=a^+}_{z=a^-}
=\lim_{\epsilon \rightarrow 0}\big(\mathbf{V}_i(z=a+\epsilon)-\mathbf{V}_i(z=a-\epsilon)\big), \epsilon >0 $  and $\mathbf{V}_{i}\big|_{z=a}=\mathbf{V}_{i}(z=a)$, for any vector $\mathbf{V}$.

The continuity conditions, (\ref{GaussB-BC}) and (\ref{Faraday-BC}), imply
that the right hand sides of equations (\ref{GaussE-BC}) and (\ref{Ampere-BC}%
) are well defined and they represent surface charge and current densities,
respectively. An immediate consequence of the boundary conditions is that
the presence of a magnetic field crossing the surface $\Sigma $ is
sufficient to generate an electric field, even in the absence of free
electric charges. For a $\theta $ boundary characterizing different $\theta $
values of the electromagnetic vacuum, these issues together with some of its
consequences over the propagation of electromagnetic waves across such
interface where studied in Refs. \cite{Huerta:2012ks,Huerta:2014ula}. It is
worth mentioning that with the modified boundary conditions, several properties of conductors in static fields still hold as far as the conductor
does not lie in the $\Sigma $ boundary. In particular, conductors are
equipotential surfaces and the electric field just outside the conductor is
normal to its surface.

\subsection{A possibility to circumvent Earnshaw's theorem}

\label{Mod_Earnshaw}

It is a well established fact in ordinary electrostatics that due to
electrostatic forces alone a charge can not be in stable equilibrium
(Earnshaw's theorem). More generally, no static object
comprised of electric charges, magnets and masses can be held
in static equilibrium by any combination of electric, magnetic or gravitational 
forces.

It is natural  to ask whether in 
the case of $\theta$ boundaries a charge can or can not be
held in stable equilibrium due to electrostatic forces alone.  This 
question is certainly relevant from an experimental point of view, but also
theoretically. As we will show below, the
case of $\theta$ ED is an interesting one as it opens a window  
to circumvent
Earnshaw's theorem. In usual electromagnetism the lack of existence of 
stable equilibrium points is proven by contradiction, owing to the fact that, 
away from the sources,
the electrostatic potential satisfies Laplace's equation.
We will show that in $\theta$ electromagnetism there exist points 
where such 
contradiction no longer takes place. From the Eq.~(\ref%
{GaussE}) and $\mathbf{E} = - \nabla \phi $ we have 
\begin{equation}
-\nabla^2 \phi = 4 \pi \rho + \tilde \theta \delta(\Sigma) \mathbf{B} \cdot 
\mathbf{n}.  \label{mod_Laplace}
\end{equation}
We inquire whether the potential has an extremum at a point 
$P$ where a test
charge is to be held in equilibrium. The argument proceeds in the usual
manner. Suppose that there exists a point $P$ at which $\phi$ acquires 
a minimum. Then at every point 
over the surface of any small closed surface enclosing $P$, one must have 
$\partial_n \phi > 0$, where $n$ denotes  the outward normal direction to the surface and thus 
one must have
\begin{equation}
\oint _{\partial \mathcal{V}} \partial_n \phi \, ds > 0.
\label{potential_extremum}
\end{equation}
where the volume $\mathcal{V}$ encloses $P$. Due to $\partial_n \phi =
\nabla \phi \cdot \mathbf{n}$ and to the divergence theorem, this latter
integral can be cast into a volume integral of the Laplacian of the
potential. Again, to isolate the contribution due to the 
$\theta$-term we set 
$\rho = 0$ to obtain 
\begin{equation}
\oint _{\partial \mathcal{V}} \partial_n \phi ds = \int_\mathcal{V} \nabla^2
\phi dv = - \tilde \theta \int_\mathcal{V} \delta(\Sigma) \mathbf{B} \cdot 
\mathbf{n} dv = - \tilde \theta \int_\Sigma \mathbf{B}_{n}\big|_{\Sigma} d^2 x.  \label{Earnshaw}
\end{equation}
In ordinary Maxwell electrostatics ($\tilde{\theta} = 0$) the right-hand-side of 
Eq. (\ref{Earnshaw}) vanishes
leading to a contradiction with Eq. (\ref{potential_extremum}) and thus proving 
that $P$ cannot be a local minimum. 
In the case of $\theta$ ED however, as far as $\tilde \theta \neq 0$ and a 
non-vanishing normal magnetic
field at $\Sigma$ exists, 
the aforementioned contradiction no longer occurs, for points $P$ at 
$\Sigma$, thus removing 
the immediate obstruction of ordinary electromagnetism for the existence of local minima. 
We emphasize that by no means 
we are claiming to have proven the existence of points of stable equilibrium in $\theta$ electrostatics, 
which can be dealt with elsewhere. To this end, 
a more thorough analysis is required along the lines of Ref. \cite{Berry:1997gm} 
in the context of ordinary
electromagnetism. We consider this an important issue. Charged or  magnetized matter can 
be held in static equilibrium in a given experimental setup. However, if this were 
possible only with the
aid of external devices as tweezers or dynamical mechanisms or EM traps \textit{e.g.},
Penning traps \cite{Penning},  these could compromise the
precision of the measurements or interact with the EM fields under study. 

\subsection{The Stress-Energy tensor}

\label{ST}

For the case of $\theta $ ED we still expect the
electromagnetic field to be an energy-momentum transmitting entity . 
In this way, for example, we should be able to compute the net force over a
given region in 3-space in terms of the values of the
electromagnetic fields on the bounding surface of that region. This
requires to establish conservation laws,  as in ordinary Maxwell
electrodynamics. In section \ref{FEandBC} we showed that the $\theta$ term
modifies the behavior of the fields at the surface $\Sigma $ only. This
suggests that, for each region in the bulk, the energy-momentum
tensor for $\theta $ ED has the same form as that in standard
electrodynamics, where the $\tilde{\theta}$ dependence
appears through the contribution to the fields arising from the additional
sources present on the boundary $\Sigma $. In fact, the identification of
the stress-energy tensor proceeds along the standard lines of
electrodynamics in a medium (see for example Ref. \cite{SCHWINGER}), where
we read the rate at which the electric field does work on the free charges 
\begin{equation}
\mathbf{J\cdot E}=-\mathbf{\nabla \cdot }\left( \frac{1}{4\pi }\mathbf{%
E\times H}\right) -\frac{1}{4\pi }\left( \mathbf{E\cdot }\frac{\partial 
\mathbf{D}}{\partial t}+\mathbf{H\cdot }\frac{\partial \mathbf{B}}{\partial t%
}\right)  \label{ECONS}
\end{equation}%
and the rate at which momentum is transferred to the charges 
\begin{equation}
\rho \mathbf{E}+\mathbf{J\times B}=-\frac{1}{4\pi }\frac{\partial }{\partial
t}\left( \mathbf{D\times B}\right) -\frac{1}{4\pi }\left[ D_{i}\mathbf{%
\nabla }E_{i}-\mathbf{\nabla }\cdot \left( \mathbf{D}\mathbf{E}\right) %
\right] -\frac{1}{4\pi }\left[ B_{i}\mathbf{\nabla }H_{i}-\mathbf{\nabla }%
\cdot \left( \mathbf{B}\mathbf{H}\right) \right] .  \label{MOMCONS}
\end{equation}%
Here the notation is\ $A_{i} \mathbf{\nabla} B_{i}=\left(
A_{i}\partial _{k}B_{i}\right) \hat{e}_{k}$ and $\mathbf{\nabla }\cdot
\left( \mathbf{AB}\right) =\left( \partial _{i}A_{i}\right) B_{k}\hat{e}_{k}$ 
for any vectors $\mathbf{A}$ and $\mathbf{B}$. Using
the constitutive relations in Eq. (\ref{CONST_REL}), we recognize from Eq. (%
\ref{ECONS}) the energy flux $\mathbf{S}$ and the energy density $U$ as 
\begin{equation}
\mathbf{S}=\frac{1}{4\pi }\mathbf{E\times B}\;\;\;\;,\;\;\;\;U=\frac{1}{8\pi 
}(\mathbf{E}^{2}+\mathbf{B}^{2}),
\end{equation}%
while from Eq. (\ref{MOMCONS}) we obtain the momentum density $\mathbf{G}$
and we identify the stress tensor $T_{ij}$ as, 
\begin{equation}
\mathbf{G}=\frac{1}{4\pi }\mathbf{E\times B}\;\;\;\;,\;\;\;\;T_{ij}=\frac{1}{%
8\pi }(\mathbf{E}^{2}+\mathbf{B}^{2})\delta _{ij}-\frac{1}{4\pi }%
(E_{i}E_{j}+B_{i}B_{j}).
\end{equation}%
Outside the free sources, the conservation equations reads 
\begin{equation}
\mathbf{\nabla \cdot S+}\frac{\partial U}{\partial t}=0\;\;\;\;,\;\;\;\;%
\frac{\partial G_{k}}{\partial t}+\partial _{i}T_{ik}=\frac{\alpha }{\pi }%
\left( \mathbf{E}\cdot \mathbf{B}\right) \partial _{k} \theta (\textbf{x}).
\end{equation}
In other words, the stress-energy tensor has in fact the same form
as in vacuum, but, as expected, it is not conserved on the $\theta $%
 boundary because of the self-induced charge and current densities arising
there.

In the standard case, the knowledge of both the stress-energy
tensor together with the GF of the system is also relevant 
in the quantum situation. For example,  when dealing with the calculation of Casimir energies, the vacuum
expectation value of the energy-momentum can be obtained with the
aid of the corresponding GF for Maxwell electrodynamics \cite{Mostepanenko:1988bs,Brown:1969na,Sopova:2005sx}. Using this method, insights regarding corrections to the Casimir energy in the context of $\theta $ ED have been developed in Refs. \cite{Canfora:2011fd,
Nie:2013}, as well as Ref. \cite{Kharlanov:2009pv} for a related perspective. This approach serves as an additional
motivation for the following  section \ref{Green}.

\section{The Green's matrix method}

\label{Green}

In this section we use the GF method to solve boundary-value problems in $%
\theta $ ED in terms of the electromagnetic potential $A^{\mu }$. Certainly
one could solve for the electric and magnetic fields from the modified
Maxwell equations (\ref{GaussE}) through (\ref{Faraday}) together with the
boundary conditions in Eqs. (\ref{GaussE-BC}-\ref{Faraday-BC}), however,
just as in ordinary electrodynamics, there might be occasions where, 
besides the knowledge of the external  sources, we are provided
with requirements on the fields at the given boundaries.
In these cases, the GF method provides the general solution, to a given
boundary value problem (Dirichlet or Neumann) for arbitrary sources. The
importance of such a general solution is evident from an experimental point
of view. For example, it would allow, at least in principle, to 
predict the electromagnetic response of topological insulators with planar,
spherical or cylindrical boundaries in the presence of more
intricate configuration of sources. Moreover, as already mentioned
in section \ref{ST}, the GF method is useful for computing the vacuum
expectation value of the energy-momentum tensor in the context of
Casimir forces. Furthermore, the GF method  should be also useful 
for the solution of dynamical problems in $\theta $ ED.

In the following we concentrate only on the static case. Since
the homogeneous Maxwell equations that express the relationship between
potentials and fields are not modified in $\theta $ ED, the electrostatic
and magnetostatic fields can be written in terms of the four-potential $%
A^{\mu }=\left( \phi ,\mathbf{A}\right) $ according to $\mathbf{E}=-\nabla
\phi $ and $\mathbf{B}=\nabla \times \mathbf{A}$, as usual. In the Coulomb
gauge $\nabla \cdot \mathbf{A}=0$, the four-potential satisfies the
equations of motion 
\begin{equation}
\left[ -\eta _{\phantom{\mu}\nu }^{\mu }\nabla ^{2}-\tilde{\theta}\delta \left( \Sigma
\right) n_{\alpha }\epsilon^{\alpha \mu \beta }_{\phantom{\alpha \mu \beta}\nu}\partial
_{\beta }\right] A^{\nu }=4\pi j^{\mu },  \label{FieldEqs2}
\end{equation}%
together with the boundary conditions (BC) 
\begin{equation}
\Delta A^{\mu }\big|_{\Sigma }=0\;\;\;\ ,\;\;\;\ \Delta \left( n^{\alpha
}\partial _{\alpha }A^{\mu }\right) \big|_{\Sigma }=-\tilde{\theta}n_{\alpha
}\epsilon^{\alpha \mu \beta }_{\phantom{\alpha \mu \beta}\nu}\partial _{\beta }A^{\nu }%
\big|_{\Sigma }.  \label{BCPotentials}
\end{equation}%
Here $n_{\alpha }$ is the unit normal to $\Sigma $, which depends on
the geometry of the $\theta $ boundary. One can further verify that
the BC of Eq. (\ref{BCPotentials}) yield those obtained in Eqs. (\ref%
{Ampere-BC}-\ref{Faraday-BC}), starting from the modified Maxwell
equations.

To obtain a general solution for the potentials $\phi $ and $\mathbf{A}$ in
the presence of arbitrary external sources $j^{\mu }\left( \mathbf{x}\right) 
$, we introduce the GF matrix $G_{\phantom{\mu}\nu }^{\mu }\left( \mathbf{x},\mathbf{x}%
^{\prime }\right) $ solving Eq. (\ref{FieldEqs2}) for a point-like source, 
\begin{equation}
\left[ -\eta _{\phantom{\mu}\nu }^{\mu }\nabla ^{2}-\tilde{\theta}\delta \left( \Sigma
\right) n_{\alpha }\epsilon^{\alpha \mu \beta }_{\phantom{\alpha \mu \beta}\nu}\partial
_{\beta }\right] G_{\phantom{\nu} \sigma }^{\nu }\left( \mathbf{x},\mathbf{x}^{\prime
}\right) =4\pi \eta _{\phantom{\mu}\sigma }^{\mu }\delta ^{3}\left( \mathbf{x}-\mathbf{x%
}^{\prime }\right) ,  \label{GF-Eq}
\end{equation}%
together with the BC of Eq. (\ref{BCPotentials}). In the following we discuss the
general solution to Eq. (\ref{GF-Eq}).  To this end we require 
an appropriate adaptation
of the standard Green's theorem, from which the
solution of Eq. (\ref{GF-Eq}) can be constructed using well-known
methods.

\subsection{Green's theorem and boundary conditions on $\Sigma$}

\label{GreenTheorem}

We begin this section by introducing the differential operator 
\begin{equation}
\mathcal{O}_{\phantom{\mu}\nu }^{\mu \phantom{\nu}i}=\eta _{\phantom{\mu}\nu
}^{\mu }\partial ^{i}+\tilde{\theta}\delta (\Sigma )n_{j}\epsilon _{%
\phantom{j\mu}\nu }^{j\mu \phantom{\nu}i},  \label{Otensor}
\end{equation}%
from which the relations  $\mathcal{O}_{\phantom{\mu}\nu }^{\mu %
\phantom{\nu}i}\partial _{i}A^{\nu }=4\pi j^{\mu }$ and $\mathcal{O}_{%
\phantom{\mu}\nu }^{\mu \phantom{\nu}i}\partial_{i}G_{\phantom{\nu}\sigma
}^{\nu }=4\pi \eta _{\phantom{\mu}\sigma }^{\mu }\delta ^{3}(\mathbf{x}-%
\mathbf{x^{\prime }})$ 
correctly yield the differential equations for the
four-potential in Eq. (\ref{FieldEqs2}) and the 
 GF in Eq. (\ref{GF-Eq}). Here the 
indexes $i,j$ range from $1$ to $3$, while $\mu ,\nu $ range from $0$ to $3$. 
These
choices reflect that we are working with static fields and that the
normal to the $\theta $ boundary is always space-like.

The relevant Green's theorem can be given in terms of two arbitrary
fields, $X_{\mu }$ and $Z_{\phantom{\mu}\nu }^{\mu }$. Defining the
tensor $T_{\phantom{\alpha}\sigma }^{i}=X_{\mu }\mathcal{O}_{\phantom{\mu}%
\nu }^{\mu \phantom{\nu}i}Z_{\phantom{\nu}\sigma }^{\nu }$ , using the
divergence theorem for $\partial _{i}T_{\phantom{\alpha}\sigma }^{i}$ and
subtracting the equation arising from the interchange $%
X\leftrightarrow Z$ we find 
\begin{eqnarray}
\oint_{S}dS\mathfrak{n}_{i}\left( X_{\mu }\mathcal{O}_{\phantom{\mu}\nu
}^{\mu \phantom{\nu}i}Z_{\phantom{\nu}\sigma }^{\nu }-Z_{\phantom{\nu}\sigma
}^{\nu }\mathcal{O}_{\phantom{\mu}\nu }^{\mu \phantom{\nu}i}X_{\mu }\right)
&=&\int_{V}d^{3}x\left[ X_{\mu }\mathcal{O}_{\phantom{\mu}\nu }^{\mu %
\phantom{\nu}i}(\partial _{i}Z_{\phantom{\nu}\sigma }^{\nu })-Z_{%
\phantom{\nu}\sigma }^{\nu }\mathcal{O}_{\phantom{\mu}\nu }^{\mu %
\phantom{\nu}i}(\partial _{i}X_{\mu })\right]  \notag  \label{GenGtheo2} \\
&&-\int_{V}d^{3}x\left[ (\partial _{i}X_{\mu })\mathcal{O}_{\phantom{\mu}\nu
}^{\mu \phantom{\nu}i}Z_{\phantom{\nu}\sigma }^{\nu }-(\partial _{i}Z_{%
\phantom{\nu}\sigma }^{\nu })\mathcal{O}_{\phantom{\mu}\nu }^{\mu %
\phantom{\nu}i}X_{\mu }\right] ,
\end{eqnarray}%
where $\mathfrak{n}$ is the outward normal to the surface $S$ bounding the
volume $V$. In deriving Eq. (\ref{GenGtheo2}) we use the result $\partial
_{i}\mathcal{O}_{\phantom{\mu}\nu }^{\mu \phantom{\nu}i}=\mathcal{O}_{%
\phantom{\mu}\nu}^{\mu \phantom{\nu}i}\partial _{i}$, which follows
directly from Eq. (\ref{Otensor}).

Substituting Eqs. (\ref{FieldEqs2}) and (\ref{GF-Eq}) in Eq. (\ref{GenGtheo2}%
), we find that the general solution for the 4-potential in the Coulomb
gauge is 
\begin{eqnarray}
A^{\mu }(\mathbf{x}) &=&\int_{V}d^{3}x^{\prime }G^{\mu}_{\phantom{\mu}\nu}(\mathbf{x},%
\mathbf{x}^{\prime })j^{\nu }(\mathbf{x}^{\prime })+\frac{1}{4\pi }%
\oint_{S}dS^{\prime }\mathfrak{n}_{i}\left[ A_{\nu }(\mathbf{x}^{\prime
})\partial ^{i}G^{\mu \nu }(\mathbf{x},\mathbf{x}^{\prime })-G^{\mu \nu }(%
\mathbf{x},\mathbf{x}^{\prime })\partial ^{i}A_{\nu }(\mathbf{x}^{\prime })%
\right]  \notag  \label{GreenMatrix} \\
&&+\frac{\tilde{\theta}}{4\pi }\int_{\Sigma }d^{2}\mathbf{x}_{\Sigma
}^{\prime }n_{j}\epsilon _{\phantom{z\mu}\nu}^{j\alpha \phantom{\nu}i}%
\left[ A_{\alpha }(\mathbf{x}^{\prime })\partial _{i}G^{\mu \nu }(\mathbf{x},%
\mathbf{x}^{\prime })-G^{\mu \nu }(\mathbf{x},\mathbf{x}^{\prime })\partial
_{i}A_{\alpha }(\mathbf{x}^{\prime })\right] _{\Sigma },
\end{eqnarray}%
where $n_{j}$ is the normal to the $\theta $ boundary. This result
yields the standard interpretation where the first term is the contribution
from the sources inside the volume $V$, the second term represents
the effects of the bounding surface $S$, while the remaining term
replaces the contributions from the surface $S$ by those of the $\theta$ boundary.

We  next  consider the issue of  the  appropriate BC for
the fields at the $\theta $ interface, when we take $S$ as a surface at
infinity where the usual BC are imposed. Inspection of Eq. (\ref{GreenMatrix}) reveals that there are four classes of BC on $\Sigma $ that specify a
solution.

The class BC-I is defined by fixing, on $\Sigma $, the scalar potential $A^{0}$ and the vector potential parallel to the the $\theta $ boundary $\mathbf{n} \times \mathbf{A}$, together with the requirement $n_{j}\epsilon_{\phantom{j\alpha}\nu}^{j\alpha \phantom{\nu}i}G^{\mu \nu }\big|_{\Sigma } = 0$ on $\Sigma$. 
This class describes the case which corresponds to
the nearest analogy with the standard Dirichlet BC. Besides, it is the 
only class for which the explicit GF is independent of the area of
the $\theta $ boundary. In section \ref{applications} we solve the problem
of a point-like charge near a planar $\theta $ boundary at fixed potential
using class BC-I.

The class BC-II is specified by fixing on $\Sigma $ the normal
component of the magnetic field $\mathbf{B}_{n}$ and the parallel component
of the electric field $\mathbf{n}\times \mathbf{E}$, plus the condition $n_{j}\epsilon_{\phantom{j \alpha}\nu }^{j\alpha \phantom{\nu}i}\partial _{i}G^{\mu \nu }\big|_{\Sigma }=1/A_{\theta }$, where $A_{\theta }=\int d^{2}%
\mathbf{x}_{\Sigma }$ is the surface area of the $\theta $ boundary. This
class corresponds to the Neumann BC in standard electrostatics, which
incorporates a factor of the inverse surface area, generating a term in the
solution involving the average contribution of the potential.

The class BC-III fixes the scalar potential $A^{0}$ and the normal
component of the magnetic field $\mathbf{B}_{n}$ on $\Sigma$, together
with the conditions $n_{j}\epsilon_{\phantom{j0}k}^{j0\phantom{k}i}  G^{\mu k}|_{\Sigma }=0$
and $n_{j}\epsilon_{\phantom{jk}0}^{jk\phantom{0}i}\partial_iG^{\mu 0}|_{\Sigma
}=1/A_{\theta }$.

The class BC-IV requires specifying $\mathbf{n}\times \mathbf{A}$ and $%
\mathbf{n}\times \mathbf{E}$ on $\Sigma $. For this class we must also
demand $n_{j}\epsilon_{\phantom{jk}0}^{jk\phantom{0}i}G^{\mu 0}|_{\Sigma }=0$ and $%
n_{j}\epsilon_{\phantom{j0}k}^{j0\phantom{k}i}\partial _{i}G^{\mu k}|_{\Sigma
}=1/A_{\theta }$.

In the next subsections we deal with the problem of constructing the GFs
for different geometrical configurations of the $\theta $ boundary, 
considering those which could be more relevant in experimental works, 
namely: planar, spherical and cylindrical $\theta $ interfaces. Moreover, having set the surface $S$  at infinity, with standard BC
there, we restrict ourselves to the case where we do not specify any
additional condition for the fields at the $\theta $ boundary. In this way, 
$A^{\mu }$ is given only by the first term of the right hand side in Eq. (%
\ref{GreenMatrix}).

\subsection{Planar $\protect\theta$ boundary}

\label{plane_Green}

\begin{figure}[tbp]
\begin{center}
\includegraphics
{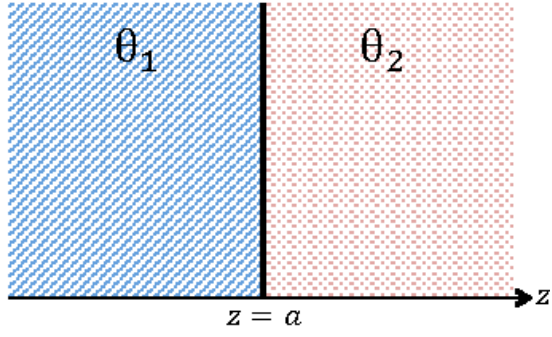}
\end{center}
\caption{{\protect\small Geometry of the semi-infinite planar $\protect\theta$-region.}}
\label{planeregions}
\end{figure}

For  the case of planar symmetry 
we choose the simplest  setup
in which the value of $\theta $ jumps across the plane $\Sigma $ and remains
constant at either side of $\Sigma $, defined by $z=a$ and indicated in Fig. %
\ref{planeregions}. In this way, the adapted coordinates to this system are
the Cartesian ones. The GF we consider has translational invariance in the
directions parallel to $\Sigma $, that is in the transverse directions $x$
and $y$, but this invariance is broken in the direction $z$. Exploiting this
symmetry we further introduce the Fourier transform in the direction
parallel to the plane $\Sigma $, taking the coordinate dependence to be $
( \mathbf{x}-\mathbf{x}^{\prime } ) _{\parallel } = (x-x^{\prime
}, y-y^{\prime })$ and define 
\begin{equation}
G_{\phantom{\mu}\nu }^{\mu }\left( \mathbf{x},\mathbf{x}^{\prime }\right) =4\pi \int 
\frac{d^{2}\mathbf{p}}{\left( 2\pi \right) ^{2}}e^{i\mathbf{p}\cdot ( 
\mathbf{x}-\mathbf{x}^{\prime } ) _{\parallel }}g_{\phantom{\mu}\nu }^{\mu }(
z,z^{\prime },\mathbf{p} ) ,  \label{RedGreenDef}
\end{equation}%
where $\mathbf{p}=(p_{x},p_{y})$ is the momentum parallel to the plane $%
\Sigma $. In the following we suppress the dependence on $\mathbf{p}
$ of the reduced GF $g_{\phantom{\mu}\nu }^{\mu }$. In this case the reduced GF
satisfies 
\begin{equation}
\left[ \partial ^{2}\eta _{\phantom{\mu} \nu }^{\mu }+i\tilde{\theta}\delta \left(
z-a\right) \epsilon _{\phantom{3\mu\alpha}\nu }^{3\mu \alpha }p_{\alpha }\right]
g_{\phantom{\nu}\sigma }^{\nu }\left( z,z^{\prime }\right) =\eta _{\phantom{\mu}\sigma }^{\mu
}\delta \left( z-z^{\prime }\right) .  \label{RedGreenFunc}
\end{equation}%
where $\partial ^{2}=\mathbf{p}^{2}-\partial _{z}^{2}$, $p^{\alpha }=\left(
0,\mathbf{p}\right) $ and $\mathbf{p}^{2}=-p^{\alpha }p_{\alpha
}=p_{x}^{2}+p_{y}^{2}$. The solution of Eq. (\ref{RedGreenFunc}) is a
simple, but not 
straight-forward task. For solving it we employ a method
similar to that used for obtaining the GF for the one-dimensional $\delta $%
-function potential in quantum-mechanics, where the free-particle GF is used
for integrating the GF equation with the $\delta $-interaction. A
main simplification arises in this case because what normally results in an
integral equation, reduces to an algebraic equation in virtue of the fact
that the integration over the $\delta $-potential can be performed. 
To proceed in an analogous way, we consider the reduced
free GF having the form $\mathcal{G}_{\phantom{\mu}\nu }^{\mu }\left( z,z^{\prime
}\right) =\mathfrak{g}\left(z,z^{\prime }\right) \eta _{\phantom{\mu}\nu }^{\mu }$,
which solves the equation
\begin{equation}
\partial ^{2}\mathcal{G}_{\phantom{\mu}\nu }^{\mu }\left( z,z^{\prime }\right) =\eta
_{\phantom{\mu}\nu }^{\mu }\delta \left( z-z^{\prime }\right) .
\end{equation}%
The details of the calculation for solving Eq. (\ref{RedGreenFunc}) were
presented in Ref. \cite{Martin-Ruiz:2015skg}. For completeness we remind here the general
solution 
\begin{equation}
g_{\phantom{\mu}\nu }^{\mu }\left( z,z^{\prime }\right) =\eta _{\phantom{\mu}\nu }^{\mu }%
\mathfrak{g}\left( z,z^{\prime }\right) -\tilde{\theta}\frac{\mathfrak{g}%
\left( z,a\right) \mathfrak{g}\left( a,z^{\prime }\right) }{1+p^{2}\tilde{%
\theta}^{2}\mathfrak{g}^{2}\left( a,a\right) } \left\lbrace \tilde{\theta}\mathfrak{%
g}\left( a,a\right) \left[ p^{\mu }p_{\nu }+\left( \eta_{\phantom{\mu}\nu }^{\mu
}+n^{\mu }n_{\nu }\right) p^{2} \right] + i \epsilon^{\mu \phantom{\nu} \alpha 3}_{\phantom{\mu}\nu} p_{\alpha } \right\rbrace ,  \label{GenSolGreenPlaneConf}
\end{equation}%
where $n _{\mu }=\left( 0,0,0,1\right) $ is the normal to $\Sigma $.

The reciprocity between the position of the unit charge and the position at
which the GF is evaluated $G_{\mu \nu }(\mathbf{x}{},{}\mathbf{x}^{\prime
})=G_{\nu \mu }(\mathbf{x}{}^{\prime },\mathbf{x}{})$ is one of its most
remarkable properties. From Eq. (\ref{RedGreenDef}) this condition requires 
\begin{equation}
g_{\mu \nu }\left( z,z^{\prime },\mathbf{p}\right) =g_{\nu \mu }\left(
z^{\prime },z,-\mathbf{p}\right) ,
\end{equation}%
which we verify directly from Eq. (\ref{GenSolGreenPlaneConf}). The symmetry 
$g_{\mu \nu }\left( z,z^{\prime }\right) =g_{\nu \mu }^{\ast }\left(
z,z^{\prime }\right) =g_{\mu \nu }^{\dagger }\left( z,z^{\prime }\right) $
is also manifest.
The various components of the static GF matrix in coordinate representation
were obtained by Fourier transforming the reduced GF, as defined in Eq. (\ref%
{RedGreenDef}). In vacuum, with no additional boundaries, the reduced GF is $%
\mathfrak{g}\left( z,z^{\prime }\right) =e^{-p|z-z^{\prime }|}/2p$, and the
corresponding GF matrix in coordinate representation is (see Ref. \cite{Martin-Ruiz:2015skg})
\begin{eqnarray}
G_{\phantom{0}0}^{0}\left( \mathbf{x},\mathbf{x}^{\prime }\right) &=&\frac{1}{|%
\mathbf{x}-\mathbf{x}^{\prime }|}-\frac{\tilde{\theta}^{2}}{4+\tilde{\theta}%
^{2}}\frac{1}{\sqrt{R^{2}+Z^{2}}},  \label{G00P} \\
G_{\phantom{0}i}^{0}\left( \mathbf{x},\mathbf{x}^{\prime }\right) &=& - \frac{2 \tilde{\theta}}{4+\tilde{\theta}^{2}} \frac{\epsilon _{0ij3} R ^{j}}{R ^{2}}\left( 1-\frac{Z}{\sqrt{R^{2}+Z^{2}}}\right) ,  \label{G0iP} \\ 
G _{\phantom{i}j}^{ i }\left( \mathbf{x},\mathbf{x}^{\prime }\right) &=& 
\eta ^{i} _{\phantom{i} j} G _{\phantom{0}0}^{ 0}\left( \mathbf{x},\mathbf{x}^{\prime }\right) - \frac{i}{2}\frac{\tilde{\theta}^{2}}{4+ \tilde{\theta}^{2}} \partial ^{i} K _{j} \left( \mathbf{x},\mathbf{x}^{\prime }\right) , \label{Gij}
\end{eqnarray}
where $Z=| z - a | + | z^{\prime } - a |$, $R ^{j} = \left( \mathbf{x-x}^{\prime }
\right) _{\parallel } ^{j}  = \left( x - x^{\prime } , y - y^{\prime} \right)$,
$R=|\left(  \mathbf{x-x}^{\prime }\right) _{\parallel }|$ and%
\begin{equation}
K ^{j} \left( \mathbf{x},\mathbf{x}^{\prime }\right) =2i\frac{\sqrt{%
R^{2}+Z^{2}}-Z}{R^{2}} R ^{j} .
\end{equation}
Finally, we observe that Eqs. (\ref{G00P}-\ref{Gij}) contain all the
required elements of the GF matrix, according to the choices of $z$ and $%
z^{\prime }$ in the function $Z$.

\subsection{Spherical $\protect\theta$ boundary}

In the preceding section, the problem of an arbitrary charge and current
distributions in the presence of a plane $\theta $ boundary was discussed by
the method of GF matrix. In this section,  following the same
procedure as in the planar situation, we discuss the spherical case, in
which the value of $\theta $ has a discontinuity across the surface $r=a$. 
 In the adapted spherical coordinates  $r,\vartheta ,\varphi $,  it proves convenient to introduce explicitly the angular
momentum operator $\hat{\mathbf{L}}=\frac{1}{i}\mathbf{x}\times \nabla $. 
 In fact, the GF equation (\ref{GF-Eq}) can be written as 
\begin{equation}
\left[ -\eta _{\phantom{\mu}\alpha }^{\mu }\nabla ^{2}-i\frac{\tilde{\theta}}{a}\delta
\left( r-a\right) \left( \eta _{\phantom{\mu}0}^{\mu }\eta _{\phantom{k}\alpha }^{k}-\eta^{\mu
k}\eta _{\phantom{0}\alpha }^{0}\right) \hat{\mathbf{L}}_{k}\right] G_{\phantom{\alpha}\nu
}^{\alpha }\left( \mathbf{x},\mathbf{x}^{\prime }\right) =4\pi \eta _{\phantom{\mu}\nu
}^{\mu }\delta ^{3}\left( \mathbf{x}-\mathbf{x}^{\prime }\right) ,
\label{GreenEqSpherical}
\end{equation}%
with $k=1,2,3$. Since the square of the angular momentum $\hat{\mathbf{L}}%
^{2}$ commutes with the operator appearing in the left hand side of Eq. (\ref%
{GreenEqSpherical}), its solution has the form 
\begin{equation}
G_{\phantom{\mu}\nu }^{\mu }\left( \mathbf{x},\mathbf{x}^{\prime }\right) =4\pi
\sum_{l=0}^{\infty }\sum_{m=-l}^{+l}\sum_{m^{\prime }=-l}^{+l}g_{lmm^{\prime
},\nu }^{\mu }\left( r,r^{\prime }\right) Y_{lm}\left( \vartheta ,\varphi
\right) Y_{lm^{\prime }}^{\ast }\left( \vartheta ^{\prime },\varphi ^{\prime
}\right) ,  \label{GreenSphericalTheta}
\end{equation}%
with the reduced GF $g_{lmm^{\prime },\nu }^{\mu }\left( r,r^{\prime
}\right) $ satisfying the equation 
\begin{equation}
\hat{\mathcal{O}}_{r}g_{lmm^{\prime },\nu }^{\mu }\left( r,r^{\prime
}\right) =\eta _{\phantom{\mu}\nu }^{\mu }\frac{\delta (r-r^{\prime })}{r^{2}}\delta
_{mm^{\prime }}+i\frac{\tilde{\theta}}{a}\delta \left( r-a\right) \left(
\eta _{\phantom{\mu}0}^{\mu }\eta _{\phantom{k}\alpha }^{k}-\eta^{\mu k}\eta _{\phantom{0}\alpha
}^{0}\right) \sum_{m^{\prime \prime }=-l}^{+l}\langle lm|\hat{\mathbf{L}}%
_{k}|lm^{\prime \prime }\rangle g_{lm^{\prime \prime }m^{\prime },\nu
}^{\alpha }\left( r,r^{\prime }\right) ,  \label{RedGreenEqSpherical2}
\end{equation}%
where $\hat{\mathcal{O}}_{r}=l\left( l+1\right) r^{-2}-r^{-2}\partial
_{r}\left( r^{2}\partial _{r}\right) $. This equation can be integrated in
the same way as for the planar symmetry. The detailed calculation is
presented in Appendix \ref{spher_Green}. The solution is 
\begin{eqnarray}
g_{lmm^{\prime },\nu }^{\mu }\left( r,r^{\prime }\right) =\eta _{\phantom{\mu}\nu
}^{\mu }\mathfrak{g}_{l}\left( r,r^{\prime }\right) \delta _{mm^{\prime
}} & - & a^{2}\tilde{\theta}^{2}l\left( l+1\right) \mathfrak{g}_{l}\left(
a,a\right) S_{l}\left( r,r^{\prime }\right) \left\langle lm\right| \hat{%
\mathbf{L}}_{\mu }\hat{\mathbf{L}}_{\nu }\left| lm^{\prime }\right\rangle 
\notag  \label{gRedComp} \\
&+& ia\tilde{\theta}S_{l}\left( r,r^{\prime }\right) \left\langle lm\right| 
\hat{\mathbf{L}}_{\alpha }\left| lm^{\prime }\right\rangle \left(\eta
_{\phantom{\mu}0}^{\mu }\Gamma _{\phantom{\alpha}\nu }^{\alpha }+\Gamma ^{\mu \alpha }\eta _{\phantom{0}\nu
}^{0}\right) ,
\end{eqnarray}%
where $\hat{\mathbf{L}}_{0}$ is the identity operator, the operator $\Gamma
^{\mu \nu }=\eta ^{\mu \nu }-\eta _{\phantom{\mu}0}^{\mu }\eta _{\phantom{\nu}0}^{\nu }$
projects a four-vector into the three-space, and 
\begin{equation}
S_{l}\left( r,r^{\prime }\right) =\frac{\mathfrak{g}_{l}\left( r,a\right) 
\mathfrak{g}_{l}\left( a,r^{\prime }\right) }{1+a^{2}\tilde{\theta}%
^{2}l\left( l+1\right) \mathfrak{g}_{l}^{2}\left( a,a\right) }.
\label{A-func}
\end{equation}%
Here $\mathfrak{g}_{l}\left( r,r^{\prime }\right) $ is the solution of the
free reduced GF equation in the absence of the $\theta $ boundary which
satisfies Eq. (\ref{O-Op-S}).
\label{other_geos_Green}
\begin{figure}[tbp]
\begin{center}
\includegraphics{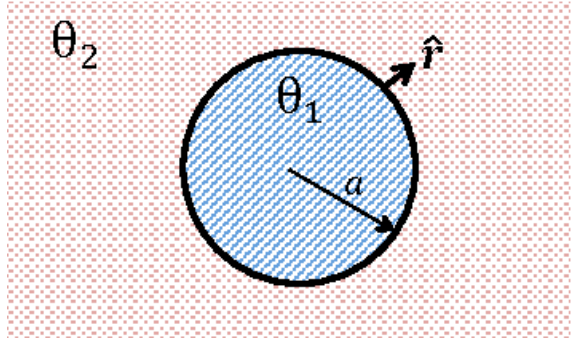}\qquad
\end{center}
\caption{{\protect\small Spherical region.}}
\label{spherical}
\end{figure}

\subsection{Cylindrical $\protect\theta$ boundary}

\begin{figure}[tbp]
\begin{center}
\includegraphics{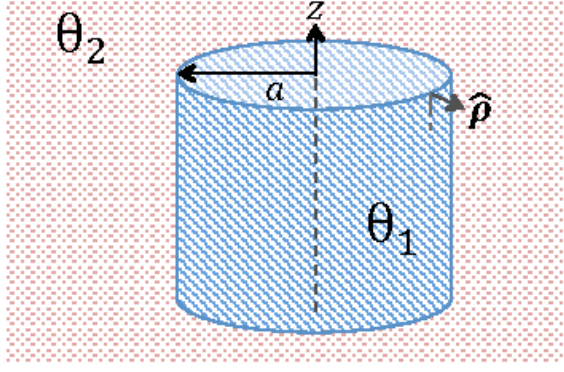}\qquad
\end{center}
\caption{{\protect\small Cylindrical region.}}
\label{cylfig}
\end{figure}

In this section we discuss the problem of an arbitrary charge and current
distributions in the presence of a cylindrical $\theta $ boundary. Let us
consider an infinite cylinder whose axis lies along the $z$-direction, such that
the value of $\theta $ has a discontinuity across its surface $\rho =a$, as
shown in Fig. \ref{cylfig}. In this way, the adapted coordinates are the
cylindrical ones: $\rho, \varphi, z$ and the GF equation is 
\begin{equation}
\left[ -\eta _{\phantom{\mu}\nu }^{\mu }\nabla^{2}-\tilde{\theta}\delta \left( \rho
-a\right) n_{\alpha }\epsilon _{\phantom{\alpha\mu\beta}\nu }^{\alpha \mu \beta }\partial
_{\beta }\right] G_{\phantom{\nu}\sigma }^{\nu }\left( \mathbf{x},\mathbf{x}^{\prime
}\right) =4\pi \eta _{\phantom{\mu}\sigma }^{\mu }\delta \left( \mathbf{x}-\mathbf{x}%
^{\prime }\right) ,  \label{GreenEqCyl}
\end{equation}%
where $n_{\alpha }=\left( 0,\cos \varphi ,\sin \varphi ,0\right) $ is the
normal to the $\theta $ interface. The GF must be invariant under
translations in the $z$ direction. In accordance with this symmetry, we
start by writing the solution of Eq. (\ref{GreenEqCyl}) as 
\begin{equation}
G_{\phantom{\mu}\nu }^{\mu }\left( \mathbf{x},\mathbf{x}^{\prime }\right) =4\pi
\int_{-\infty }^{+\infty }\frac{dk}{2\pi }e^{ik\left( z-z^{\prime }\right) }%
\frac{1}{2\pi }\sum_{m=-\infty }^{+\infty }\sum_{m^{\prime }=-\infty
}^{+\infty }g_{mm^{\prime },\nu }^{\mu }\left( \rho ,\rho ^{\prime
};k\right) e^{i ( m\varphi -m^{\prime }\varphi ^{\prime } ) }.
\label{GreenFuncCyl}
\end{equation}
with the reduced GF $g_{mm^{\prime },\nu }^{\mu }\left( \rho ,\rho ^{\prime
};k\right) $ satisfying the equation 
\begin{equation}
\hat{\mathcal{O}}_{\rho }^{(m)}g_{mm^{\prime },\sigma }^{\mu }-i\tilde{\theta%
}\delta \left( \rho -a\right) \left[k\sum_{m^{\prime \prime}=-\infty
}^{+\infty }A_{m^{\prime \prime }m,\nu }^{\mu }g_{m^{\prime \prime
}m^{\prime },\sigma }^{\nu }+\epsilon _{\phantom{1\mu 2} \nu }^{1\mu 2}\,\frac{m}{\rho }%
\delta _{mm^{\prime }}g_{mm^{\prime },\sigma }^{\nu }\right] =\eta
_{\phantom{\mu}\sigma }^{\mu }\frac{\delta \left( \rho -\rho ^{\prime }\right) }{\rho }%
\delta _{mm^{\prime }},  \label{GreenEqCyl4}
\end{equation}%
where 
\begin{equation}
\hat{\mathcal{O}}_{\rho }^{(m)}=-\frac{1}{\rho }\frac{\partial }{\partial
\rho }\left( \rho \frac{\partial }{\partial \rho }\right) +\frac{m^{2}}{\rho
^{2}}+k^{2}\;\;\;\ ,\;\;\;\ A_{mm^{\prime \prime },\nu }^{\mu }=\frac{1}{2}%
\left[ \delta _{m,m^{\prime \prime }-1}{(\tilde{\epsilon}_{\phantom{\mu}\nu }^{\mu
})}^\ast+\delta _{m,m^{\prime \prime }+1}\tilde{\epsilon}_{\phantom{\mu}\nu }^{\mu }\right] , 
\end{equation}
with $\tilde{\epsilon}_{\phantom{\mu}\nu }^{\mu }=\epsilon _{\phantom{1 \mu 3} \nu }^{1\mu
3}+i\epsilon _{\phantom{2 \mu 3}\nu }^{2\mu 3}$. This equation can be integrated in
the same way as for the planar and spherical cases. Detailed calculations
are presented in Appendix \ref{cyl_Green}.

The solution is 
\begin{eqnarray}
g_{mm^{\prime },\sigma }^{0}\left( \rho ,\rho ^{\prime }\right) &=&\eta
_{\phantom{0}\sigma }^{0}\delta _{mm^{\prime }}\left[ \mathfrak{g}_{m}\left( \rho
,\rho ^{\prime }\right) -\tilde{\theta}^{2}\mathfrak{f}_{m}\left( k\right)
C_{mm}\left( \rho ,\rho ^{\prime }\right) \right] +i\tilde{\theta}\left(
m\delta _{mm^{\prime }}\eta _{\phantom{3}\sigma }^{3}+kaA_{m^{\prime }m,\sigma
}^{0}\right) C_{mm^{\prime }}\left( \rho ,\rho ^{\prime }\right) ,
\label{g0nCyl-3} \\
g_{mm^{\prime },\sigma }^{3}\left( \rho ,\rho ^{\prime }\right) &=&\eta
_{\phantom{3}\sigma }^{3}\delta _{mm^{\prime }}\left[ \mathfrak{g}_{m}\left( \rho
,\rho ^{\prime }\right) -m^{2}\tilde{\theta}^{2}\mathfrak{g}_{m}\left(
a,a\right) C_{mm}\left( \rho ,\rho ^{\prime }\right) \right] +im\tilde{\theta%
}\left( \eta_{\phantom{0}\sigma }^{0}+ika\tilde{\theta}A_{mm^{\prime },\sigma
}^{0}\right) C_{mm^{\prime }}\left( \rho ,\rho ^{\prime }\right) ,
\label{g3nCyl-2} \\
g_{mm^{\prime },j}^{i}\left( \rho ,\rho ^{\prime }\right) &=&\eta
_{\phantom{i}j}^{i}\delta _{mm^{\prime }}\mathfrak{g}_{m}\left( \rho ,\rho ^{\prime
}\right) -\tilde{\theta}^{2}k^{2}a^{2}\mathfrak{g}_{m}\left( \rho ,a\right)
\sum_{m^{\prime \prime }=-\infty }^{+\infty }A_{m^{\prime
}m,0}^{j}A_{m^{\prime }m^{\prime \prime },j}^{0}C_{m^{\prime \prime
}m^{\prime }}\left( a,\rho ^{\prime }\right) ,  \label{ginCyl-2}
\end{eqnarray}%
where $\mathfrak{f}_{m}\left( k\right) =m^{2}\mathfrak{g}_{m}\left(
a,a\right) +\frac{k^{2}a^{2}}{2}\left[ \mathfrak{g}_{m+1}\left( a,a\right) +%
\mathfrak{g}_{m-1}\left( a,a\right) \right] $ and 
\begin{equation}
C_{mm^{\prime }}\left( \rho ,\rho ^{\prime }\right) =\frac{\mathfrak{g}%
_{m}\left( \rho ,a\right) \mathfrak{g}_{m^{\prime }}\left( a,\rho ^{\prime
}\right) }{1+\tilde{\theta}^{2}\mathfrak{f}_{m}\left( k\right) \mathfrak{g}%
_{m}\left( a,a\right) }.  \label{C-function}
\end{equation}%
Here $\mathfrak{g}_{m^{\prime }}\left( \rho ,\rho ^{\prime }\right) $ is the
solution to the reduced GF equation (\ref{GreenEqCyl4}) in the absence of
the $\theta $ boundary. Note that the remaining $\mu \nu -$components of the
GF matrix can be obtained from the symmetry property 
\begin{equation}
g_{mm^{\prime}, \mu \nu }\left( \rho ,\rho ^{\prime };k\right) =g_{m^{\prime
}m,\nu \mu }\left( \rho ^{\prime },\rho ;-k\right) .  \label{SymGFCyl}
\end{equation}

\section{Applications}

\label{applications}

\subsection{Point-like charge near a planar $\protect\theta$ boundary at
fixed potentials}

Now we deal with the problem of one or more point charges in the presence of
a planar $\theta $ boundary surface at fixed potential. This case falls
under the BC-I class of boundary conditions at the $\theta $ interface. In
the planar case these BC reduce to demanding  the full BC-I Green
function ${(G_{\rm I})}_{\phantom{\mu}\nu }^{\mu }\left( \mathbf{x},\mathbf{x}^{\prime }\right) $ to
be zero. As an example let us consider a point-like electric charge in front
of an infinite planar $\theta $ boundary at zero potentials $(A^{0},\mathbf{A%
}_{\parallel })=(0,\mathbf{0})$ located at $z=0$. No additional bounding
surfaces are considered. As discussed in Ref. \cite{Martin-Ruiz:2015skg} the problem of a
point-like charge in front of an infinite planar $\theta $ boundary is
equivalent to the problem of the original charge, together with an electric
charge and a magnetic monopole located at the mirror-image point behind the
plane but with the $\theta$ interface removed \textit{i.e.,} $\tilde \theta =0$. 
These electric and magnetic images reproduce the boundary conditions
in Refs. (\ref{GaussE-BC}-\ref{Faraday-BC}), induced by the non-trivial jump
of the $\theta $-value.

The problem at hand can also be solved in terms of images.  Following
similar steps as in the standard case,  the corresponding GF, ${(G_{\rm I})}_{\phantom{\mu}\nu }^{\mu }\left( \mathbf{x},\mathbf{x}^{\prime }\right) $, can be constructed
as 

\begin{equation}
{(G_{\rm I})}_{\phantom{\mu}\nu }^{\mu }\left( \mathbf{x},\mathbf{x}^{\prime }\right)= G_{\phantom{\mu}\nu}^{\mu }\left( \mathbf{x},\mathbf{x}^{\prime }\right) +F_{\phantom{\mu}\nu }^{\mu
}\left( \mathbf{x},\mathbf{x}^{\prime }\right) ,  \label{DGF}
\end{equation}%

with the addition of a matrix $F_{\phantom{\mu} \nu }^{\mu }$ which satisfies the
homogeneous equation 
\begin{equation}
\left[ -\eta _{\phantom{\mu}\nu }^{\mu }\nabla ^{2}-\tilde{\theta}\delta \left(
z\right) \epsilon _{\phantom{3 \mu\alpha}\nu }^{3\mu \alpha }\partial _{\alpha }\right]
F_{\phantom{\nu}\sigma }^{\nu }\left( \mathbf{x},\mathbf{x}^{\prime }\right) =0.
\label{DGF2}
\end{equation}%
This freedom in the definition of the GF allow us to choose appropriately
the matrix $F_{\phantom{\mu}\nu }^{\mu }$ in such a way that ${(G_{\rm I})}_{\phantom{\mu}\nu }^{\mu }\left( 
\mathbf{x},\mathbf{x}^{\prime }\right) =0$ for $\mathbf{x}^{\prime }$ on $%
\Sigma $. Let us consider a point-like  electric charge located at $%
z^{\prime }>0$, and assume that $z>0$. The required components of the GF
matrix for solving this problem are 
\begin{eqnarray}
G_{\phantom{0}0}^{0}\left( \mathbf{x},\mathbf{x}^{\prime }\right) &=&\frac{1}{|%
\mathbf{x}-\mathbf{x}^{\prime }|}-\frac{\tilde{\theta}^{2}}{4+\tilde{\theta}%
^{2}}\frac{1}{|\mathbf{x}-\mathbf{x}^{\prime \prime }|},  \label{G00} \\
G_{\phantom{0}i}^{0}\left( \mathbf{x},\mathbf{x}^{\prime }\right) &=&-\frac{2\tilde{%
\theta}}{4+\tilde{\theta}^{2}}\frac{\epsilon _{0ij3}\left( \mathbf{x}-%
\mathbf{x}^{\prime }\right) _{\parallel }^{j}}{R^{2}}\left( 1-\frac{%
z+z^{\prime }}{|\mathbf{x}-\mathbf{x}^{\prime \prime }|}\right) ,
\label{G0i}
\end{eqnarray}%
where $\mathbf{x}^{\prime }=(x^{\prime },y^{\prime },z^{\prime })$ denotes
the position of the charge and $\mathbf{x}^{\prime \prime }=(x^{\prime
},y^{\prime },-z^{\prime })$ indicates the position of the images.

Regarding the problem of the BC-I Green function, we find that 
\begin{eqnarray}
{(G_{\rm I})}_{\phantom{0}0}^{0}\left( \mathbf{x},\mathbf{x}^{\prime }\right) &=&G_{\phantom{0}0}^{0}\left( 
\mathbf{x},\mathbf{x}^{\prime }\right) -G_{\phantom{0}0}^{0}\left( \mathbf{x},\mathbf{%
x}^{\prime \prime }\right) =\left( 1+\frac{\tilde{\theta}^{2}}{4+\tilde{%
\theta}^{2}}\right) \left( \frac{1}{|\mathbf{x}-\mathbf{x}^{\prime }|}-\frac{%
1}{|\mathbf{x}-\mathbf{x}^{\prime \prime }|}\right) , \\
{(G_{\rm I})}_{\phantom{0}i }^{0 }\left( \mathbf{x},\mathbf{x}^{\prime }\right) &=&G_{\phantom{0}i}^{0}\left( 
\mathbf{x},\mathbf{x}^{\prime }\right) -G_{\phantom{0}i}^{0}\left( \mathbf{x},\mathbf{%
x}^{\prime \prime }\right) =\frac{2\tilde{\theta}}{4+\tilde{\theta}^{2}}%
\frac{\epsilon _{0ij3}\left( \mathbf{x}-\mathbf{x}^{\prime }\right)
_{\parallel }^{j}}{R^{2}}\left( \frac{z+z^{\prime }}{|\mathbf{x}-\mathbf{x}%
^{\prime \prime }|}-\frac{z-z^{\prime }}{|\mathbf{x}-\mathbf{x}^{\prime }|}%
\right) .
\end{eqnarray}%
which effectively vanish at $z^{\prime }=0$. We interpret the fields as
follows. The electric field in the region $z>0$ is generated by four
electric charges: (i) the original charge of 
unit strength
at $z^{\prime }$, (ii) an equal and opposite charge located at the mirror-image point, (iii)
an electric charge of strength $-\tilde{\theta}^{2}/(4+\tilde{\theta}^{2})$
located at the mirror-image point and (iv) an electric charge of strength $+%
\tilde{\theta}^{2}/(4+\tilde{\theta}^{2})$ located at $z^{\prime }$. The
magnetic field can be interpreted as that produced by two monopoles, one
of strength $2\tilde{\theta}/(4+\tilde{\theta}^{2})$ located at the
mirror-image point, induced by the $\theta $ boundary, and the other of
strength $-2\tilde{\theta}/(4+\tilde{\theta}^{2})$ at $z^{\prime }$, arising from the BC $\mathbf{A}_{\mathbf{\Vert }}|_{\Sigma }=0$.

In a similar fashion one can further check that the electric field in the
region $z<0$ is due to an electric charge of strength $-4/(4+\tilde{\theta}%
^{2})$ located at $z^{\prime }$ plus an electric charge of strength $+4/(4+%
\tilde{\theta}^{2})$ located at $-z^{\prime }$. The magnetic field can be
interpreted as being generated by two magnetic monopoles, one of strength $-2%
\tilde{\theta}/(4+\tilde{\theta}^{2})$ located at $z^{\prime }$ together
with its image of strength $+2\tilde{\theta}/(4+\tilde{\theta}^{2})$ located
at $-z^{\prime }.$

\subsection{Point-like charge near a spherical $\protect\theta$ boundary}

The problem we shall discuss is that of a point-like charge in vacuum ($%
\theta _{2} = 0 $) near a spherical topological medium of radius $a$ and $%
\theta _{1} \neq 0 $. For simplicity we choose the line connecting the
center of the sphere and the point charge as the $z$-axis. Thus the current density can be written as $j ^{\mu} \left( \mathbf{x} ^{\prime} \right) = \frac{q}{b ^{2}} \eta ^{\mu} _{\phantom{\mu}0} \delta \left( r ^{\prime} - b \right) \delta \left( \cos \vartheta
^{\prime} - 1 \right) \delta \left( \varphi ^{\prime} \right) $, with $b > a$.

The solution for this problem is then 
\begin{eqnarray}
\phi \left( \mathbf{x}\right) &=&\int G_{\phantom{0}\mu }^{0}\left( \mathbf{x},%
\mathbf{x}^{\prime }\right) j ^{\mu }\left( \mathbf{x}^{\prime }\right) d^{3}%
\mathbf{x}^{\prime }=qG_{\phantom{0}0}^{0}\left( \mathbf{x},\mathbf{b}\right) ,
\label{ScalPotSphere} \\
\mathbf{A}\left( \mathbf{x}\right) &=&\sum_{k=1}^{3}\int G_{\phantom{k}\mu
}^{k}\left( \mathbf{x},\mathbf{x}^{\prime }\right) j^{\mu }\left( \mathbf{x}%
^{\prime }\right) \hat{\mathbf{e}}_{k}d^{3}\mathbf{x}^{\prime
}=\sum_{k=1}^{3}qG_{\phantom{k}0}^{k}\left( \mathbf{x},\mathbf{b}\right) \hat{\mathbf{%
e}}_{k},  \label{VectPotSphere}
\end{eqnarray}%
where $\mathbf{b}=b\hat{\mathbf{e}}_{z}$. With the use of the corresponding
components of the GF matrix, Eq. (\ref{gRedComp}), the scalar and vector
potentials are 
\begin{eqnarray}
\phi \left( \mathbf{x}\right) &=&4\pi q\sum_{l=0}^{\infty
}\sum_{m=-l}^{+l}\sum_{m^{\prime }=-l}^{+l}\left[ \mathfrak{g}_{l}\left(
r,b\right) -a^{2}\tilde{\theta}^{2}l\left( l+1\right) \mathfrak{g}_{l}\left(
a,a\right) S_{l}\left( r,b\right) \right] \delta _{mm^{\prime }}Y_{lm}\left(
\vartheta ,\varphi \right) Y_{lm^{\prime }}^{\ast }\left( 0,\varphi ^{\prime
}\right) ,  \label{PotSphere2} \\
\mathbf{A}\left( \mathbf{x}\right) &=&4\pi q\sum_{l=0}^{\infty
}\sum_{m=-l}^{+l}\sum_{m^{\prime }=-l}^{+l}ia\tilde{\theta}S_{l}\left(
r,b\right) Y_{lm}\left( \vartheta ,\varphi \right) Y_{lm^{\prime }}^{\ast
}\left( 0,\varphi ^{\prime }\right) \left\langle lm\right| \hat{\mathbf{L}}%
\left| lm^{\prime }\right\rangle .  \label{PotVecSphere2}
\end{eqnarray}
From the relations $Y_{lm}\left( 0,\varphi \right) =\delta _{m0}\sqrt{\frac{%
2l+1}{4\pi }}$ and $Y_{l0}\left( \vartheta ,\varphi \right) =\sqrt{\frac{2l+1%
}{4\pi }}P_{l}\left( \cos \vartheta \right) $ we find 
\begin{eqnarray}
\phi \left( \mathbf{x}\right) &=&q\sum_{l=0}^{\infty }\left[ \mathfrak{g}%
_{l}\left( r,b\right) -a^{2}\tilde{\theta}^{2}l\left( l+1\right) \mathfrak{g}%
_{l}\left( a,a\right) S_{l}\left( r,b\right) \right] \left( 2l+1\right)
P_{l}\left( \cos \vartheta \right) ,  \label{PotSphere3} \\
\mathbf{A}\left( \mathbf{x}\right) &=&q\sum_{l=0}^{\infty }\sum_{m=-l}^{+l}ia%
\tilde{\theta}\sqrt{4\pi \left( 2l+1\right) }S_{l}\left( r,b\right)
Y_{lm}\left( \vartheta ,\varphi \right) \left\langle lm\right| \hat{\mathbf{L%
}}\left| l0\right\rangle .  \label{PotVecSphere3}
\end{eqnarray}%
which immediately yields $A_{z}=0$. The remaining components of the vector
potential can be calculated by introducing the combinations $A_{\pm
}=A_{x}\pm iA_{y}$. The result is 
\begin{equation}
A_{\pm }\left( \mathbf{x}\right) =q\sum_{l=0}^{\infty }ia\tilde{\theta}\sqrt{%
4\pi \left( 2l+1\right) l\left( l+1\right) }S_{l}\left( r,b\right) Y_{l\pm
1}\left( \vartheta ,\varphi \right) ,  \label{PotVecSphere4pm}
\end{equation}%
where the expected symmetry $A_{+}=A_{-}^{\ast }$ follows from the relation $%
Y_{l1}^{\ast }=-Y_{l-1}$. Recalling that 
\begin{equation}
A_{\theta }=-\sin \vartheta A_{z}+\frac{1}{2}\cos \vartheta \left(
A_{+}e^{-i\varphi }+A_{-}e^{i\varphi }\right),  \quad  A_{\varphi }=\frac{1}{%
2i}\left( A_{+}e^{-i\varphi }-A_{-}e^{i\varphi }\right) ,
\end{equation}%
we obtain 
\begin{equation}
A_{\theta }=0\;\;\;\ ,\;\;\;\ A_{\varphi }=q\sum_{l=0}^{\infty }a\tilde{%
\theta}\left( 2l+1\right) S_{l}\left( r,b\right) \frac{\partial P_{l}\left(
\cos \vartheta \right) }{\partial \vartheta },  \label{AFIN}
\end{equation}%
in spherical coordinates.

Now we analyze the field strengths for the regions (1): $r>b>a$ and (2): $%
b>a>r$. Using the free GF, $\mathfrak{g}_{l}\left( r,r^{\prime }\right) =%
\frac{1}{2l+1}\frac{r_{<}^{l}}{r_{>}^{l+1}}$, the scalar potential and the (nonzero component of the) vector potential for the region (1) take the form 
\begin{eqnarray}
\phi _{(1)}\left( \mathbf{x}\right) &=&\frac{q}{|\mathbf{x}-\mathbf{b}|}%
-q\sum_{l=1}^{\infty }\frac{\tilde{\theta}^{2}l\left( l+1\right) }{\left(
2l+1\right) ^{2}+\tilde{\theta}^{2}l\left( l+1\right) }\frac{a^{2l+1}}{%
r^{l+1}b^{l+1}}P_{l}\left( \cos \gamma \right) ,  \label{PotSphere5} \\
A_{(1)\varphi }\left( \mathbf{x}\right) &=&q\sum_{l=0}^{\infty }\frac{\tilde{%
\theta}\left( 2l+1\right) }{\left( 2l+1\right) ^{2}+\tilde{\theta}%
^{2}l\left( l+1\right) }\frac{a^{2l+1}}{r^{l+1}b^{l+1}}\frac{\partial
P_{l}\left( \cos \vartheta \right) }{\partial \vartheta },
\label{PotVecSphere5}
\end{eqnarray}%
respectively. The corresponding electric and magnetic field strengths can be
calculated directly
from
$\mathbf{E}=-\nabla \phi $ and $\mathbf{B}=\nabla
\times \mathbf{A}$, respectively. The result is 
\begin{eqnarray}
\mathbf{E}_{(1)}\left( \mathbf{x}\right) &=&q\frac{\mathbf{x}-\mathbf{b}}{|%
\mathbf{x}-\mathbf{b}|^{3}}-q\sum_{l=1}^{\infty }\frac{\tilde{\theta}%
^{2}l\left( l+1\right) }{\left( 2l+1\right) ^{2}+\tilde{\theta}^{2}l\left(
l+1\right) }\frac{a^{2l+1}}{r^{l+2}b^{l+1}}\left[ -\left( l+1\right)
P_{l}\left( \cos \vartheta \right) \hat{\mathbf{r}}+\frac{\partial
P_{l}\left( \cos \vartheta \right) }{\partial \vartheta }\hat{\vartheta}%
\right] ,  \label{ElectFieldSphere} \\
\mathbf{B}_{(1)}\left( \mathbf{x}\right) &=&q\sum_{l=1}^{\infty }\frac{\tilde{%
\theta}l\left( 2l+1\right) }{\left( 2l+1\right) ^{2}+\tilde{\theta}%
^{2}l\left( l+1\right) }\frac{a^{2l+1}}{r^{l+2}b^{l+1}}\left[ -\left(
l+1\right) P_{l}\left( \cos \vartheta \right) \hat{\mathbf{r}}+\frac{%
\partial P_{l}\left( \cos \vartheta \right) }{\partial \vartheta }\hat{%
\vartheta}\right] .  \label{MagFieldSphere}
\end{eqnarray}%
We now ask what is the behavior of these field strengths when the separation
between the point charge and the sphere is large compared to the radius of
the sphere, $b\gg a$. Since the $l$-th term in the sum behaves as $\left(
a/b\right) ^{l+1}$, only small values of $l$ contribute. The leading
contribution arises from $l=1$ and we obtain 
\begin{eqnarray}
\mathbf{E}_{(1)}\left( \mathbf{x}\right) &\sim &q\frac{\mathbf{x}-\mathbf{b}}{|%
\mathbf{x}-\mathbf{b}|^{3}}+\frac{p}{r^{3}}\left( 2\cos \vartheta \hat{%
\mathbf{r}}+\sin \vartheta \hat{\vartheta}\right) ,
\label{ElectFieldSphere2} \\
\mathbf{B}_{(1)}\left( \mathbf{x}\right) &\sim &\frac{m}{r^{3}}\left( 2\cos
\vartheta \hat{\mathbf{r}}+\sin \vartheta \hat{\vartheta}\right) ,
\label{MagFieldSphere2}
\end{eqnarray}%
which corresponds to the electric and magnetic fields generated by an
electric dipole $\mathbf{p}$ and a magnetic dipole $\mathbf{m}$ lying at the
origin and pointing in the $z$ direction 
\begin{equation*}
\mathbf{p}=\frac{2q\tilde{\theta}}{9+2\tilde{\theta}^{2}}\frac{a^{3}}{b^{2}}%
\hat{\mathbf{e}}_{z}=-\frac{2}{3}\mathbf{m}.
\end{equation*}

Next we consider the field strengths in the region (2): $b>a>r$. The scalar
potential and the  $\varphi $-component of the vector potential become 
\begin{eqnarray}
\phi _{(2)}\left( \mathbf{x}\right) &=&\frac{q}{|\mathbf{x}-\mathbf{b}|}%
-q\sum_{l=1}^{\infty }\frac{\tilde{\theta}^{2}l\left( l+1\right) }{\left(
2l+1\right) ^{2}+\tilde{\theta}^{2}l\left( l+1\right) }\frac{r^{l}}{b^{l+1}}%
P_{l}\left( \cos \gamma \right) ,  \label{PotSphere6} \\
A_{(2)\varphi }\left( \mathbf{x}\right) &=&q\sum_{l=0}^{\infty }\frac{\tilde{%
\theta}\left( 2l+1\right) }{\left( 2l+1\right) ^{2}+\tilde{\theta}%
^{2}l\left( l+1\right) }\frac{r^{l}}{b^{l+1}}\frac{\partial P_{l}\left( \cos
\vartheta \right) }{\partial \vartheta },  \label{PotVecSphere6}
\end{eqnarray}%
respectively. The corresponding field strengths are 
\begin{eqnarray}
\mathbf{E}_{(2)}\left( \mathbf{x}\right) &=&q\frac{\mathbf{x}-\mathbf{b}}{|%
\mathbf{x}-\mathbf{b}|^{3}}+q\sum_{l=1}^{\infty }\frac{\tilde{\theta}%
^{2}l\left( l+1\right) }{\left( 2l+1\right) ^{2}+\tilde{\theta}^{2}l\left(
l+1\right) }\frac{r^{l-1}}{b^{l+1}}\left[ lP_{l}\left( \cos \vartheta
\right) \hat{\mathbf{r}}+\frac{\partial P_{l}\left( \cos \vartheta \right) }{%
\partial \vartheta }\hat{\vartheta}\right] ,  \label{ElectFieldSphere3} \\
\mathbf{B}_{(2)}\left( \mathbf{x}\right) &=&-q\sum_{l=1}^{\infty }\frac{%
\tilde{\theta}l\left( 2l+1\right) \left( l+1\right) }{\left( 2l+1\right)
^{2}+\tilde{\theta}^{2}l\left( l+1\right) }\frac{r^{l-1}}{b^{l+1}}\left[
lP_{l}\left( \cos \vartheta \right) \hat{\mathbf{r}}+\frac{\partial
P_{l}\left( \cos \vartheta \right) }{\partial \vartheta }\hat{\vartheta}%
\right] .  \label{MagFieldSphere3}
\end{eqnarray}%
When the separation between the point charge and the sphere is large
compared to the radius of the sphere, $b\gg a$, the field strengths in the
region $r<a$ become 
\begin{eqnarray}
\mathbf{E}_{(2)}\left( \mathbf{x}\right) &\sim &q\frac{\mathbf{x}-\mathbf{b}}{%
|\mathbf{x}-\mathbf{b}|^{3}}-\frac{1}{3}\mathbf{P},
\label{ElectFieldSphere4} \\
\mathbf{B}_{(2)}\left( \mathbf{x}\right) &\sim &\frac{2}{3}\mathbf{M},
\label{MagFieldSphere4}
\end{eqnarray}%
where 
\begin{equation}
\mathbf{P}=\frac{6q\tilde{\theta}}{9+2\tilde{\theta}^{2}}\frac{1}{b^{2}}\hat{%
\mathbf{e}}_{z}=-\frac{2}{3}\mathbf{M}.
\end{equation}%
An interesting feature to note is the form of the field strengths in such
region. The electric field behaves as the field produced by a uniformly
polarized sphere with polarization $\mathbf{P}$, while the magnetic field
resembles the one produced by a uniformly magnetized sphere with
magnetization $\mathbf{M}$.

\subsection{Infinitely straight current-carrying wire near a cylindrical $%
\protect\theta$ boundary}

Let us consider an infinite straight wire parallel to the $z$-axis carrying
a current $I$ in the $+z$ direction. The wire is located in vacuum ($\theta
_{2}=0$) at a distance $b$ from the $z$-axis. Also we assume a cylindrical
non-trivial topological insulator of radius $a<b$, with its axis parallel to the wire and passing through the origin. For simplicity we choose
the coordinates such that $\varphi ^{\prime }=0$. Therefore the current
density is $j^{\mu }\left( \mathbf{x}^{\prime }\right) =\frac{I}{b}\eta
_{\phantom{\mu}3}^{\mu }\delta \left( \varphi ^{\prime }\right) \delta \left( \rho
^{\prime }-b\right) $. The solution for this problem is 
\begin{equation}
A^{\mu }\left( \mathbf{x}\right) =\int G_{\phantom{\mu}3}^{\mu }\left( \mathbf{x},%
\mathbf{x}^{\prime }\right) j^{3}\left( \mathbf{x}^{\prime }\right) d\mathbf{%
x}^{\prime }=2I\lim_{k\rightarrow 0}\sum_{m=-\infty }^{+\infty
}\sum_{m^{\prime }=-\infty }^{+\infty }g_{mm^{\prime },3}^{\mu }\left( \rho
,b,k\right) e^{im\varphi },  \label{4-Pot-Cyl}
\end{equation}%
where the various components of the reduced GF in cylindrical coordinates
are given by Eqs. (\ref{g0nCyl-3}-\ref{ginCyl-2}). With the use of the
corresponding components of the GF matrix, the scalar and the (nonzero
component of the) vector potential are 
\begin{eqnarray}
A^{0}\left( \mathbf{x}\right)  &=&-4I\tilde{\theta}\lim_{k\rightarrow
0}\sum_{m=1}^{+\infty }mC_{mm}\left( \rho ,b\right) \sin m\varphi ,
\label{SPot-Cyl} \\
A^{3}\left( \mathbf{x}\right)  &=&2I\lim_{k\rightarrow 0}\left\{ \mathfrak{g}%
_{0}\left( \rho ,b\right) +2\sum_{m=1}^{+\infty }\left[ \mathfrak{g}%
_{m}\left( \rho ,b\right) -\tilde{\theta}^{2}m^{2}\mathfrak{g}_{m}\left(
a,a\right) C_{mm}\left( \rho ,b\right) \right] \cos m\varphi \right\} .
\label{VPot-Cyl}
\end{eqnarray}
The reduced GF in vacuum is $\mathfrak{g}_{m}\left( \rho ,\rho ^{\prime
};k\right) =\mbox{I}_{m}\left( k\rho _{<}\right) \mbox{K}_{m}\left( k\rho
_{>}\right) $, where $\mbox{I}_{m}$ and $\mbox{K}_{m}$ are the modified
Bessel functions of the first and second kind respectively. Using the
limiting form of the modified Bessel functions for small arguments \cite%
{Gradshteyn}, the limit $k\rightarrow 0$ of the reduced GF becomes 
\begin{equation}
\lim_{k\rightarrow 0}\mathfrak{g}_{m}\left( \rho ,\rho ^{\prime };k\right) =%
\frac{1}{2m}\left( \frac{\rho _{<}}{\rho _{>}}\right) ^{m},
\end{equation}%
and then $\lim_{k\rightarrow 0}\mathfrak{f}_{m}\left( k\right) =m/2$. The
potentials now become 
\begin{eqnarray}
A^{0}\left( \mathbf{x}\right)  &=&-\frac{4I\tilde{\theta}}{4+\tilde{\theta}%
^{2}}\sum_{m=1}^{+\infty }\frac{1}{m}\left( \frac{a_{<}}{\rho _{>}}\frac{a}{b%
}\right) ^{m}\sin m\varphi ,  \label{SPot-Cyl2} \\
A^{3}\left( \mathbf{x}\right)  &=&2I\left\{ -\log \rho
_{>}+\sum_{m=1}^{+\infty }\left[ \frac{1}{m}\left( \frac{b_{<}}{\rho _{>}}%
\right) ^{m}-\frac{\tilde{\theta}^{2}}{4+\tilde{\theta}^{2}}\frac{1}{m}%
\left( \frac{a_{<}}{\rho _{>}}\frac{a}{b}\right) ^{m}\right] \cos m\varphi
\right\} ,  \label{VPot-Cyl2}
\end{eqnarray}%
where the symbols $>$ and $<$ denotes the greater and lesser in the ratio $%
a/\rho $. The summations can be performed analytically, with the result 
\begin{eqnarray}
A^{0}\left( \mathbf{x}\right)  &=&\frac{4I\tilde{\theta}}{4+\tilde{\theta}%
^{2}}\arctan \left( \frac{\sin \varphi }{\cos \varphi -\frac{\rho _{>}}{a_{<}%
}\frac{b}{a}}\right) ,  \label{SPot-Cyl3} \\
A^{3}\left( \mathbf{x}\right)  &=&I\left\{ -\log \left[ \rho ^{2}-2b\rho
\cos \varphi +b^{2}\right] +\frac{\tilde{\theta}^{2}}{4+\tilde{\theta}^{2}}%
\log \left[ 1-2\frac{a_{<}}{\rho _{>}}\frac{a}{b}\cos \varphi +\left( \frac{%
a_{<}}{\rho _{>}}\frac{a}{b}\right) ^{2}\right] \right\} .  \label{VPot-Cyl3}
\end{eqnarray}%
Next we analyze the field strengths for the regions (1): $\rho >b>a$ and (2): 
$b>a>\rho $. 

In the region (1) the potentials take the form 
\begin{eqnarray}
A_{(1)}^{0}\left(\mathbf{x}\right)  &=&\frac{4I\tilde{\theta}}{4+\tilde{\theta%
}^{2}}\arctan \left( \frac{\sin \varphi }{\cos \varphi -\frac{\rho }{d}}%
\right) .  \label{SPot-Cyl4} \\
A_{(1)}^{3}\left( \mathbf{x}\right)  &=&I\left[ -\log \left( \rho
^{2}+b^{2}-2b\rho \cos \varphi \right) +\frac{\tilde{\theta}^{2}}{4+\tilde{%
\theta}^{2}}\log \left( 1-2\frac{d}{\rho }\cos \varphi +\frac{d^{2}}{\rho
^{2}}\right) \right] ,  \label{VPot-Cyl4}
\end{eqnarray}%
where $d=a^{2}/b$. The corresponding electric and magnetic field can be
calculated directly as $\mathbf{E}=-\nabla A^{0}$ and $\mathbf{B}=\nabla
\times \mathbf{A}$, respectively. The result is 
\begin{eqnarray}
\mathbf{E}_{(1)}\left( \mathbf{x}\right)  &=&-\frac{4I\tilde{\theta}}{4+\tilde{%
\theta}^{2}}\left[ \frac{d\sin \varphi }{\rho ^{2}+d^{2}-2d\rho \cos \varphi 
}\hat{\rho}+\left( \frac{1}{\rho }-\frac{\rho -d\cos \varphi }{\rho
^{2}+d^{2}-2d\rho \cos \varphi }\right) \hat{\varphi}\right] ,
\label{EF-Cyl} \\
\mathbf{B}_{(1)}\left( \mathbf{x}\right)  &=&\hat{\rho}\left[ \frac{-2Ib\sin
\varphi }{\rho ^{2}+b^{2}-2b\rho \cos \varphi }+\frac{2I\tilde{\theta}^{2}}{%
4+\tilde{\theta}^{2}}\frac{d\sin \varphi }{\rho ^{2}+d^{2}-2d\rho \cos
\varphi }\right]   \notag \\
&&+\hat{\varphi}\left[ \frac{2I\left( \rho -b\cos \varphi \right) }{\rho
^{2}+b^{2}-2b\rho \cos \varphi }+\frac{2I\tilde{\theta}^{2}}{4+\tilde{\theta}%
^{2}}\left( \frac{1}{\rho }-\frac{\rho -d\cos \varphi }{\rho
^{2}+d^{2}-2d\rho \cos \varphi }\right) \right] .  \notag  \label{MF-Cyl}
\end{eqnarray}%
The fields can be interpreted as follows. The magnetic field corresponds to
that generated by the wire with current $I$, plus  two image currents: one
of strength $I\tilde{\theta}^{2}/(4+\tilde{\theta}^{2})$ located at the
origin, and the other of strength $-I\tilde{\theta}^{2}/(4+\tilde{\theta}%
^{2})$ located at $d=a^{2}/b$. 

In the region (2) the potentials are 
\begin{eqnarray}
A_{(2)}^{0}\left(\mathbf{x}\right)  &=&\frac{4I\tilde{\theta}}{4+\tilde{%
\theta}^{2}}\arctan \left( \frac{\sin \varphi }{\cos \varphi -\frac{b}{\rho }%
}\right) ,  \label{SPot-Cyl5} \\
A_{(2)}^{3}\left(\mathbf{x}\right)  &=&\frac{4I}{4+\tilde{\theta}^{2}}\log
\left( \rho ^{2}+b^{2}-2b\rho \cos \varphi \right) .  \label{VPot-Cyl5}
\end{eqnarray}%
The fields can be calculated directly. The result is 
\begin{eqnarray}
\mathbf{E}_{(2)}\left(\mathbf{x}\right)  &=&\frac{4I\tilde{\theta}}{4+\tilde{%
\theta}^{2}}\left[ \frac{b\sin \varphi }{\rho ^{2}+b^{2}-2b\rho \cos \varphi 
}\hat{\rho}-\frac{\rho -b\cos \varphi }{\rho ^{2}+b^{2}-2b\rho \cos \varphi }%
\hat{\varphi}\right] ,  \label{SPot-Cyl6} \\
\mathbf{B}_{(2)}\left(\mathbf{x}\right)  &=&\frac{4I}{4+\tilde{\theta}^{2}}%
\left[ -\frac{2b\sin \varphi }{\rho ^{2}+b^{2}-2b\rho \cos \varphi }\hat{\rho%
}+\frac{2\left( \rho -b\cos \varphi \right) }{\rho ^{2}+b^{2}-2b\rho \cos
\varphi }\hat{\varphi}\right] .  \label{VPot-Cyl6}
\end{eqnarray}%
The magnetic field in this region corresponds to the one produced by a
current $4I/(4+\tilde{\theta}^{2})$ located at $b$.

\subsection{Infinitely uniformly charged wire near a cylindrical $\protect%
\theta$ boundary}

Let us consider an infinite straight wire which carries the uniform charge
per unit length $\lambda $. The wire is placed parallel to the $z$-axis and
is located in vacuum ($\theta _{2}=0$) at a distance $b$ from the $z$-axis.
Also we assume a cylindrical non-trivial topological insulator of radius $a<b
$. Again,  we choose the coordinates such that $\varphi ^{\prime }=0$.
Therefore the current density is $j^{\mu }\left( \mathbf{x}^{\prime }\right)
=\frac{\lambda }{b} \eta ^{\mu} _{\phantom{\mu}0} \delta \left( \varphi ^{\prime }\right) \delta \left(
\rho ^{\prime }-b\right)$. The solution for this problem is 
\begin{equation}
A^{\mu }\left( \mathbf{x}\right) =\int G_{\phantom{\mu}0}^{\mu }\left( \mathbf{x},%
\mathbf{x}^{\prime }\right) j^{0}\left( \mathbf{x}^{\prime }\right) d\mathbf{%
x}^{\prime }=2\lambda \lim_{k\rightarrow 0}\sum_{m=-\infty }^{+\infty
}\sum_{m^{\prime }=-\infty }^{+\infty }g_{mm^{\prime },0}^{\mu }\left( \rho
,b,k\right) e^{im\varphi }.  \label{4-Pot-Cyl-UC}
\end{equation}%
The nonzero components can be calculated in the same way as in the previous
example. The final result is 
\begin{eqnarray}
A^{0}\left( \mathbf{x}\right)  &=&\lambda \left\{ -\log \left[ \rho
^{2}-2b\rho \cos \varphi +b^{2}\right] +\frac{\tilde{\theta}^{2}}{4+\tilde{%
\theta}^{2}}\log \left[ 1-2\frac{a_{<}}{\rho _{>}}\frac{a}{b}\cos \varphi
+\left( \frac{a_{<}}{\rho _{>}}\frac{a}{b}\right) ^{2}\right] \right\} ,
\label{SPot-Cyl-UC} \\
A^{3}\left( \mathbf{x}\right)  &=&\frac{4\lambda \tilde{\theta}}{4+\tilde{%
\theta}^{2}}\arctan \left( \frac{\sin \varphi }{\cos \varphi -\frac{\rho _{>}%
}{a_{<}}\frac{b}{a}}\right) .  \label{VPot-Cyl-UC}
\end{eqnarray}%
Now we analyze the field strengths for the regions (1): $\rho>b>a$ and (2): $%
b>a>\rho $. In the region (1) the potentials take the form 
\begin{eqnarray}
A_{(1)}^{0}\left( \mathbf{x}\right)  &=&\lambda \left[ -\log \left[ \rho
^{2}-2b\rho \cos \varphi +b^{2}\right] +\frac{\tilde{\theta}^{2}}{4+\tilde{%
\theta}^{2}}\log \left( 1-2\frac{d}{\rho }\cos \varphi +\frac{d^{2}}{\rho
^{2}}\right) \right] ,  \label{SPot-Cyl4-UC} \\
A_{(1)}^{3}\left( \mathbf{x}\right)  &=&\frac{4\lambda \tilde{\theta}}{4+%
\tilde{\theta}^{2}}\arctan \left( \frac{\sin \varphi }{\cos \varphi -\frac{%
\rho }{d}}\right) ,  \label{VPot-Cyl4-UC}
\end{eqnarray}%
where $d=a^{2}/b$. The corresponding electric and magnetic fields can be
calculated as usual. In the region (1) the result is 
\begin{eqnarray}
\mathbf{E}_{(1)}\left( \mathbf{x}\right)  &=&\left[ \frac{2\lambda \left( \rho
-b\cos \varphi \right) }{\rho ^{2}+b^{2}-2\rho b\cos \varphi }+\frac{%
2\lambda \tilde{\theta}^{2}}{4+\tilde{\theta}^{2}}\left( \frac{1}{\rho }-%
\frac{\rho -d\cos \varphi }{\rho ^{2}+d^{2}-2d\rho \cos \varphi }\right) %
\right] \hat{\rho}  \notag \\
&&+\left( \frac{2\lambda b\sin \varphi }{\rho ^{2}+b^{2}-2\rho b\cos \varphi 
}-\frac{2\lambda \tilde{\theta}^{2}}{4+\tilde{\theta}^{2}}\frac{d\sin
\varphi }{\rho ^{2}+d^{2}-2d\rho \cos \varphi }\right) \hat{\varphi}, \\
\mathbf{B}_{(1)}\left( \mathbf{x}\right)  &=&\frac{4\lambda \tilde{\theta}}{4+%
\tilde{\theta}^{2}}\left[ \left( \frac{1}{\rho }-\frac{\rho -d\cos \varphi }{%
\rho ^{2}+d^{2}-2d\rho \cos \varphi }\right) \hat{\rho}-\frac{d\sin \varphi 
}{\rho ^{2}+d^{2}-2d\rho \cos \varphi }\hat{\varphi}\right] .
\end{eqnarray}%
In the region (2) the result is 
\begin{eqnarray}
\mathbf{E}_{(2)}\left( \mathbf{x}\right)  &=&\frac{4\lambda }{4+\tilde{\theta}%
^{2}}\left[ \frac{2\left( \rho -b\cos \varphi \right) }{\rho
^{2}+b^{2}-2\rho b\cos \varphi }\hat{\rho}+\frac{2b\sin \varphi }{\rho
^{2}+b^{2}-2\rho b\cos \varphi }\hat{\varphi}\right] , \\
\mathbf{B}_{(2)}\left( \mathbf{x}\right)  &=&\frac{4\lambda \tilde{\theta}}{4+%
\tilde{\theta}^{2}}\left( \hat{\rho}\frac{\rho -b\cos \varphi }{\rho
^{2}+b^{2}-2b\rho \cos \varphi }+\hat{\varphi}\frac{b\sin \varphi }{\rho
^{2}+b^{2}-2b\rho \cos \varphi }\right) .
\end{eqnarray}


\section{Summary }

\label{summary}

In this paper we have considered an appealing topological extension of
Maxwell electrodynamics which constitutes a low energy effective theory to study the response of topological
insulators. 
The model is defined by supplementing the Lagrange density of classical
electrodynamics in $3+1$ spacetime dimensions  with the
$U(1)$
Pontryagin invariant coupled to a scalar field $\theta$. 
We take the
field $\theta$ as an external prescribed quantity that is
function of 
space.
This  coupling
violates Lorentz, parity and time-reversal symmetries, while
preserving gauge invariance. We have restricted to the case where 
$\theta$ is
piecewise constant in two different regions of space separated by a common
interface $\Sigma$, denoted also as the $\theta$ boundary.  Nevertheless,  our methods can be directly generalized to include additional 
spatial
interfaces. 
The related problem of considering $n_\mu$
of time-like nature is interesting on its own, but it lies out of the scope of this work and can be dealt with elsewhere. This would model a system with 
$\theta = \theta (t)$ rather than $\theta(\textbf{x})$.  One can anticipate that to tackle this problem correctly, a fully dynamical theory would be necessary. 
The $\theta$-value can be thought of as an effective parameter
characterizing the properties of a novel electromagnetic media, possibly arising
from a more fundamental theory of matter, which encodes the effect of novel 
quantum degrees of freedom. 
We have
referred to this model as $\theta$-electrodynamics 
$\theta$ ED.
In this scenario the field equations in the bulk remain the standard Maxwell
equations but the discontinuity of $\theta$ at the surface $\Sigma$ alters the behavior of the
fields, as shown in Eq.(\ref{FieldEqs}), giving rise to magnetoelectric effects
such as a nontrivial 
Faraday-
and Kerr-like rotation of the plane of
polarization of electromagnetic waves traversing the interface $\Sigma$ 
as analyzed in \cite{Hehl-Obukov, Huerta:2012ks,Huerta:2014ula}. 
In a preceding work \cite{Martin-Ruiz:2015skg}, we introduced the Green's function method in
 static $\theta$ ED and we  provided the calculation of  the Green's function
for a planar $\theta$ boundary, which is summarized  here in Eqs. (\ref{G00P}-\ref{Gij}), for completeness. As a first application, we tackled the 
problem of a point-like
electric charge located near the planar $\theta$ interface, and we  recovered the results of Ref. \cite{science} which were obtained with the use of the
 method  of images. However the Green's function approach is far more general
given that it is well suited to deal with the calculation of electric and
magnetic fields arising from arbitrary sources. The force between the charge
and the $\theta$ boundary was  also computed by two different methods: (i) we used the GF
to  calculate  the interaction energy between a charge-current distribution and
the  $\theta$ boundary, with the vacuum energy removed. Alternatively, (ii) we arrived at
the same result by considering the momentum flux perpendicular to the
interface   in terms of the stress energy tensor. The problem of an infinitely straight current-carrying wire near
a planar $\theta$ boundary was also discussed in detail.

In this work we  extend our method  to compute  the static Green's function for a $\theta$ boundary with 
spherical and cylindrical geometries, given the
fact that those  geometries seems to  be relevant for a large number of 
experimental  settings. The results are presented in 
Eq. (\ref{gRedComp})  for the spherical case and 
Eqs. (\ref{g0nCyl-3}-\ref{ginCyl-2}) for the cylindrical case. 
Prior to this we have dealt with some important structural aspects of classical 
static electromagnetic theory. Namely the issue of the possibility of
having stable equilibrium due to electromagnetic forces only and the correct
construction of the stress-energy tensor to further analyze the status of 
conservation laws. Regarding the former, in Eq. (\ref{Earnshaw}) we have
shown that for the case of $\theta$ ED, points of stable equilibrium are
not a priori forbidden, at least not in a trivial way as in ordinary electromagnetism. 
Thus TIs as modeled by $\theta$ interfaces 
have a chance  
to circumvent Earnshaw's theorem. With respect to the stress-energy tensor, 
we found that the energy density, energy flux, momentum density and the stress
tensor are defined in the usual way, however the ensuing conservation laws
reveal a non-conservation of the stress-energy tensor on the 
$\theta$ boundary, which in retrospect is not unexpected since the 
mere existence of the boundary breaks translational symmetry along the direction
perpendicular to it.
Also, as shown in Eq. (\ref{GreenMatrix}),  we  extend the  Green's theorem to {$\theta$ ED} and we
classify the boundary conditions that can be imposed on $\Sigma$ in four
different classes. Class I makes contact with the Dirichlet boundary-value
problem in standard electrostatics, while the remaining  classes yield boundary conditions depending  on the
surface area of the $\theta$ boundary.  These are the most important results of our work, since they allow to obtain  the corresponding static electric and magnetic fields for arbitrary sources and arbitrary boundary conditions in the given geometries. Also, the method provides a well defined  starting point for either analytical or  numerical approximations in the cases where the exact analytical calculations are not possible.
 As an illustration of the extended Green's theorem, we analyze the problem of a point-like electric charge near a grounded  planar $\theta$ boundary,  \textit{i.e.}, having zero scalar potential,
together with zero parallel components of the vector potential.  In this case,  the boundary conditions imply that the Green's function is  zero at the $\theta$ boundary. In close analogy with the standard case, this GF  is subsequently constructed starting from the original plane symmetric GF given in 
Eqs. (\ref{G00P}-\ref{Gij}), by adding an homogeneous solution of Eq. (\ref{GF-Eq}) in order to fulfill the boundary conditions. In this simple situation, the method of images allows  to readily identify these solutions and the final configuration can be interpreted in terms of suitable images charges and induced magnetic monopoles.
Regarding the interpretation of the solution as due to image charges and 
image magnetic monopoles, let us  recall that these images are just artifacts. In fact, the physical 
situation under study is mimicked by an hypothetical one with the same physical
sources with the
$\theta$ boundary removed plus the fictitious sources (image charges and monopoles) to ensure that the boundary conditions of the fields are met at the 
location of the boundary. The appearance of magnetic monopoles  in this
solution seems to violate the Maxwell law $\nabla \cdot \mathbf{B}=0$,
which remained unaltered in the case of $\theta$ ED. 
However, this is not the case. In fact, given  $\left( \mathbf{x}\pm \mathbf{r}%
\right) /|\mathbf{x}\pm \mathbf{r}|^{3}\sim \mathbf{\nabla }_{x}(1/|\mathbf{x%
}\pm \mathbf{r}|)$, we have $\nabla \cdot \mathbf{B\sim \nabla }_{x}^{2}(1/|%
\mathbf{x}\pm \mathbf{r}|) \sim \delta (\mathbf{x} \pm \mathbf{r})$ in a
region where $\mathbf{x\neq }\pm \mathbf{r}$. Physically, the magnetic field
is induced by a surface current density 
$\mathbf{J}=\tilde{\theta}\delta \left( z\right) \mathbf{E}\times \mathbf{n}$
that is circulating around the origin. 
Additional applications include the use of the 
spherical Green's function  to analyze the problem of a point-like
charge near a spherical $\theta$ boundary, while the cylindrical Green's
function allows for  the calculation of the fields produced by  an infinitely current carrying wire and
by a uniformly charged wire, both  near a cylindrical $\theta$ boundary an parallel to its axis. 

The Green's function method should also be useful for the
extension to the dynamical case. In this respect, to our knowledge, little
efforts have been done in the context of topological insulators.
Furthermore, Green's functions are also relevant for the computation of
other effects, such as the Casimir effect. 

The method here expanded and initiated in Ref. \cite{Martin-Ruiz:2015skg}, when applied to specific configurations 
representing given experimental setups, predicts results that coincide 
with those in  previously existing literature. Our method however, enjoys 
a certain generality in the sense that can be applied to more intricate configuration
of sources, in which case, for example, the method of images can result more cumbersome. 

\acknowledgments  LFU
acknowledges J. Zanelli for introducing him to the $\theta$-theories. 
MC has been supported by the project FONDECYT (Chile)
Initiation into Research Grant No. 11121633 and also wants to thank the kind
hospitality at Instituto de Ciencias Nucleares, UNAM. LFU has been supported
in part by the project No. IN104815 from Direcci\'on General Asuntos del
Personal Acad\'emico (Universidad Nacional Aut\'onoma de M\'exico) and the
project CONACyT (M\'exico) \# 237503. LFU and AMR thank the warm hospitality
at Universidad Andres Bello.

\appendix

\section{GF for a spherical $\protect\theta$ boundary}

\label{spher_Green}

In this section we construct the GF in spherical coordinates for the
configuration shown in  Fig. \ref{spherical} where the $\theta $ boundary is
the surface of a sphere of radius $a$ with center at the origin. Here the
adapted coordinate system is provided by spherical coordinates. The various
components of the GF are the solution of 
\begin{equation}
\left[ -\eta _{\phantom{\mu}\alpha }^{\mu }\nabla ^{2}-i\frac{\tilde{\theta}}{a}\delta
\left( r-a\right) \left( \eta _{\phantom{\mu}0}^{\mu }\eta _{\phantom{k}\alpha }^{k}-\eta^{\mu
k}\eta _{\phantom{0}\alpha }^{0}\right) \hat{\mathbf{L}}_{k}\right] G_{\phantom{\alpha}\nu
}^{\alpha }\left( \mathbf{x},\mathbf{x}^{\prime }\right) =4\pi \eta _{\phantom{\mu}\nu
}^{\mu }\delta \left( \mathbf{x}-\mathbf{x}^{\prime }\right) ,  \label{GE-S}
\end{equation}%
with $k=1,2,3$. Here $\hat{\mathbf{L}}_{k}$ are the components of the
angular momentum operator. Since the completeness relation for the spherical
harmonics is 
\begin{equation}
\delta \left( \cos \vartheta -\cos \vartheta ^{\prime }\right) \delta \left(
\varphi -\varphi ^{\prime }\right) =\sum_{l=0}^{\infty
}\sum_{m=-l}^{+l}Y_{lm}\left( \vartheta ,\varphi \right) Y_{lm}^{\ast
}\left( \vartheta ^{\prime },\varphi ^{\prime }\right) ,  \label{S-Comp}
\end{equation}%
we look for a solution of form 
\begin{equation}
G_{\phantom{\mu}\nu }^{\mu }\left( \mathbf{x},\mathbf{x}^{\prime }\right) =4\pi
\sum_{l=0}^{\infty }\sum_{l^{\prime }=0}^{\infty
}\sum_{m=-l}^{+l}\sum_{m^{\prime }=-l^{\prime }}^{+l^{\prime }}g_{ll^{\prime
}mm^{\prime },\nu }^{\mu }\left( r,r^{\prime }\right) Y_{lm}\left( \vartheta
,\varphi \right) Y_{l^{\prime }m^{\prime }}^{\ast }\left( \vartheta ^{\prime
},\varphi ^{\prime }\right) ,  \label{GF-S}
\end{equation}%
where $g_{ll^{\prime }mm^{\prime },\nu }^{\mu }\left( r,r^{\prime }\right) $
is the reduced GF analogous to $g_{\phantom{\mu} \nu }^{\mu }(z,z^{\prime })$ in the
case of planar symmetry. The operator in the left hand side of Eq. (\ref%
{GE-S}) commutes with $\mathbf{\hat{L}}^{2}$ in such a way that 
\begin{equation}
g_{ll^{\prime }mm^{\prime },\nu }^{\mu }\left( r,r^{\prime }\right) =\delta
_{ll^{\prime }}g_{lmm^{\prime },\nu }^{\mu }\left( r,r^{\prime }\right) .
\label{RGF-S}
\end{equation}

In the limiting case $\tilde{\theta}\rightarrow 0$ the matrix elements take
the simple form $g_{lm,\nu }^{\mu }\left( r,r^{\prime }\right) =\eta _{\phantom{\mu}\nu
}^{\mu }\mathfrak{g}_{l}\left( r,r^{\prime }\right) $, where $\mathfrak{g}%
_{l}\left( r,r^{\prime }\right) $ solves the equation 
\begin{equation}
\hat{\mathcal{O}}_{r}\mathfrak{g}_{l}\left( r,r^{\prime }\right) =\frac{%
\delta \left( r-r^{\prime }\right) }{r^{2}},  \label{O-Op-S}
\end{equation}%
with the radial operator being 
\begin{equation}
\hat{\mathcal{O}}_{r}=\frac{l\left( l+1\right) }{r^{2}}-\frac{1}{r^{2}}\frac{%
\partial }{\partial r}\left( r^{2}\frac{\partial }{\partial r}\right) .
\label{O-Op-S2}
\end{equation}%
The solution to Eq. (\ref{O-Op-S}) for different configurations is
well-known (see for example Ref. \cite{SCHWINGER}). In free space,with 
boundary conditions at infinity, the solution is 
\begin{equation}
\mathfrak{g}_{l}\left( r,r^{\prime }\right) =\frac{r_{<}^{l}}{r_{>}^{l+1}}%
\frac{1}{2l+1},  \label{free-RGF-S}
\end{equation}%
where $r_{>}$ ($r_{<}$) is the greater (lesser) of $r$ and $r^{\prime }$.
The substitution of Eq. (\ref{free-RGF-S}) into Eq. (\ref{GF-S}) correctly
reproduces the well-known result $|\mathbf{x}-\mathbf{x}^{\prime }|^{-1}$.

In the following we focus in determining the various components of the GF
matrix in  Eq. (\ref{GE-S}) . The method we shall employ is similar to that
used for solving the planar case, but the required mathematical techniques
are more subtle because the dependence upon the angular momentum operator.

Substituting Eq. (\ref{GF-S}) into Eq. (\ref{GE-S}) and using $-\nabla
^{2}\rightarrow \hat{\mathcal{O}}_{r}$ gives  
\begin{align}
& \sum_{l=0}^{\infty }\sum_{m=-l}^{+l}\left[ \eta _{\phantom{\mu}\alpha }^{\mu }\hat{%
\mathcal{O}}_{r}-i\frac{\tilde{\theta}}{a}\delta \left( r-a\right) \left(
\eta _{\phantom{\mu}0}^{\mu }\eta _{\phantom{k}\alpha }^{k}-\eta ^{\mu k}\eta _{\phantom{0}\alpha
}^{0}\right) \hat{\mathbf{L}}_{k}\right] g_{lmm^{\prime },\nu }^{\alpha
}\left( r,r^{\prime }\right) Y_{lm}\left( \vartheta ,\varphi \right)
Y_{lm^{\prime }}^{\ast }\left( \vartheta ^{\prime },\varphi ^{\prime
}\right)   \notag \\
& =\eta_{\phantom{\mu}\nu }^{\mu }\sum_{l=0}^{\infty }\sum_{m=-l}^{+l}Y_{lm}\left(
\vartheta ,\varphi \right) Y_{lm^{\prime }}^{\ast }\left( \vartheta ^{\prime
},\varphi ^{\prime }\right) \frac{\delta (r-r^{\prime })}{r^{2}}\delta
_{mm^{\prime }}.  \label{RGFE-S}
\end{align}%
The linear independence of $\ $the spherical harmonics $Y_{lm^{\prime
}}^{\ast }\left( \vartheta ^{\prime },\varphi ^{\prime }\right) $ yields 
\begin{equation}
\sum_{l=0}^{\infty }\sum_{m=-l}^{+l}\left[ \eta _{\phantom{\mu}\alpha }^{\mu }\hat{%
\mathcal{O}}_{r}-i\frac{\tilde{\theta}}{a}\delta \left( r-a\right) \left(
\eta _{\phantom{\mu}0}^{\mu }\eta _{\phantom{k}\alpha }^{k}-\eta ^{\mu k}\eta _{\phantom{0}\alpha
}^{0}\right) \hat{\mathbf{L}}_{k}\right] g_{lmm^{\prime },\nu }^{\alpha
}\left( r,r^{\prime }\right) Y_{lm}\left( \vartheta ,\varphi \right) =\eta
_{\phantom{\mu}\nu }^{\mu }\sum_{l=0}^{\infty }\sum_{m=-l}^{+l}Y_{lm}\left( \vartheta
,\varphi \right) \frac{\delta (r-r^{\prime })}{r^{2}}\delta _{mm^{\prime }}
\label{RGFE-S2}
\end{equation}

Next we multiply Eq. (\ref{RGFE-S2}) to the left by $Y_{l^{\prime \prime
}m^{\prime \prime }}^{\ast }(\vartheta ,\varphi )$ and integrate over the
solid angle $d\Omega (\vartheta ,\varphi )$. After using the properties of
the spherical harmonics
\begin{equation}
\langle l^{\prime \prime }m^{\prime \prime }|lm\rangle =\delta _{l^{\prime
\prime }l}\delta _{m^{\prime \prime }m}\;\;\;\;,\;\;\;\;\langle l^{\prime
\prime }m^{\prime \prime }|\hat{\mathbf{L}}_{k}|lm\rangle =\delta
_{l^{\prime \prime }l}\;\;\langle l^{\prime \prime }m^{\prime \prime }|\hat{%
\mathbf{L}}_{k}|l^{\prime \prime }m\rangle ,  \label{SH1}
\end{equation}
where 
\begin{equation}
\left\langle lm\right| \hat{\mathbf{L}}_{k}\left| lm^{\prime \prime
}\right\rangle =\int_{\Omega }Y_{lm}^{\ast }\left( \vartheta ,\varphi
\right) \hat{\mathbf{L}}_{k}Y_{lm^{\prime \prime }}\left( \vartheta ,\varphi
\right) d\Omega ,  \label{ME-S}
\end{equation}
we obtain 
\begin{equation}
\hat{\mathcal{O}}_{r}g_{lmm^{\prime },\nu }^{\mu }\left( r,r^{\prime
}\right) -\eta _{\phantom{\mu}\nu }^{\mu }\frac{\delta (r-r^{\prime })}{r^{2}}\delta
_{mm^{\prime }}=i\frac{\tilde{\theta}}{a}\delta \left( r-a\right) \left(
\eta _{\phantom{\mu}0}^{\mu }\eta _{\phantom{k}\alpha}^{k}-\eta ^{\mu k}\eta _{\phantom{0}\alpha
}^{0}\right) \sum_{m^{\prime \prime }=-l}^{+l}\langle lm|\hat{\mathbf{L}}%
_{k}|lm^{\prime \prime }\rangle g_{lm^{\prime \prime }m^{\prime },\nu
}^{\alpha }\left( r,r^{\prime }\right) ,  \label{RGFE-S3}
\end{equation}
where we have relabeled $l^{\prime \prime }\rightarrow l$ and $m^{\prime
\prime }\leftrightarrow m$.

The resulting equation can be integrated using the free
reduced GF $\mathfrak{g}_{l}\left( r,r^{\prime }\right)$, satisfying Eq. (%
\ref{O-Op-S}), with the result 
\begin{equation}
g_{lmm^{\prime },\nu }^{\mu }\left( r,r^{\prime }\right) =\eta _{\phantom{\mu}\nu
}^{\mu }\mathfrak{g}_{l}\left( r,r^{\prime }\right) \delta _{mm^{\prime }}+i
a\tilde{\theta} \left( \eta^{\mu}_{\phantom{\mu} 0}\eta_{\phantom{k}\alpha }^{k} - \eta^{\mu
k} \eta^{0} _{\phantom{0}\alpha }\right) \mathfrak{g}_{l}\left( r,a\right)
\sum_{m^{\prime \prime }=-l}^{+l}g_{lm^{\prime \prime }m^{\prime },\nu
}^{\alpha }\left( a,r^{\prime }\right) \left\langle lm\right| \hat{\mathbf{L}%
}_{k}\left| lm^{\prime \prime }\right\rangle .  \label{RGFE-S4}
\end{equation}

Now we have to solve for the various components. To this end we split
Eq. (\ref{RGFE-S4}) into the components  $\mu =0$ and $\mu =k$; 
\begin{eqnarray}
g_{lmm^{\prime },\nu }^{0}\left( r,r^{\prime }\right)  &=&\eta _{\phantom{0}\nu }^{0}%
\mathfrak{g}_{l}\left( r,r^{\prime }\right) \delta _{mm^{\prime }}+ia\tilde{%
\theta}\mathfrak{g}_{l}\left( r,a\right) \sum_{k=1}^{3}\sum_{m^{\prime
\prime }=-l}^{+l}g_{lm^{\prime \prime }m^{\prime },\nu }^{k}\left(
a,r^{\prime }\right) \left\langle lm\right| \hat{\mathbf{L}}_{k}\left|
lm^{\prime \prime }\right\rangle ,  \label{g0n-S} \\
g_{lmm^{\prime },\nu }^{k}\left( r,r^{\prime }\right)  &=&\eta _{\phantom{k}\nu }^{k}%
\mathfrak{g}_{l}\left( r,r^{\prime }\right) \delta _{mm^{\prime }}+ia\tilde{%
\theta}\mathfrak{g}_{l}\left( r,a\right) \sum_{m^{\prime \prime
}=-l}^{+l}g_{lm^{\prime \prime }m^{\prime },\nu }^{0}\left( a,r^{\prime
}\right) \left\langle lm\right| \hat{\mathbf{L}}_{k}\left| lm^{\prime \prime
}\right\rangle ,  \label{gkn-S}
\end{eqnarray}%
where the second term in the right-hand side produces the coupling between
the two types of components. Now we set $r=a$ in Eq. (\ref{gkn-S}) and then
substitute into Eq. (\ref{g0n-S}) yielding 
\begin{equation}
g_{lmm^{\prime },\nu }^{0}\left( r,r^{\prime }\right) =\eta _{\phantom{0}\nu }^{0}%
\mathfrak{g}_{l}\left( r,r^{\prime }\right) \delta _{mm^{\prime }}+ia\tilde{%
\theta}\mathfrak{g}_{l}\left( r,a\right) \mathfrak{g}_{l}\left( a,r^{\prime
}\right) \left\langle lm\right| \eta _{\phantom{k}\nu }^{k}\hat{\mathbf{L}}_{k}\left|
lm^{\prime }\right\rangle -a^{2}\tilde{\theta}^{2}l\left( l+1\right) 
\mathfrak{g}_{l}\left( a,a\right) \mathfrak{g}_{l}\left( r,a\right)
g_{lmm^{\prime },\nu }^{0}\left( a,r^{\prime }\right) ,
\label{g0n-S2}
\end{equation}%
where we have used the result 
\begin{equation}
\sum_{k=1}^{3}\sum_{m^{\prime }=-l}^{+l}\left\langle lm\right| \hat{\mathbf{L%
}}_{k}\left| lm^{\prime }\right\rangle \left\langle lm^{\prime }\right| \hat{%
\mathbf{L}}_{k}\left| lm^{\prime \prime }\right\rangle
=\sum_{k=1}^{3}\left\langle lm\right| \hat{\mathbf{L}}_{k}^{2}\left|
lm^{\prime \prime }\right\rangle =l\left( l+1\right) \delta _{mm^{\prime
\prime }}.  \label{SH2}
\end{equation}%
Now we set  $r=a$ in Eq. (\ref{g0n-S2})  and solve it for $g_{lmm^{\prime
},\nu }^{0}\left( a,r^{\prime }\right) $, obtaining  
\begin{equation}
g_{lmm^{\prime },\nu }^{0}\left( a,r^{\prime }\right) =\frac{\eta _{\phantom{0}\nu
}^{0}\delta _{mm^{\prime }}+ia\tilde{\theta}\mathfrak{g}_{l}\left(
a,a\right) \sum_{k=1}^{3}\left\langle lm\right| \eta _{\phantom{k}\nu }^{k}\hat{\mathbf{%
L}}_{k}\left| lm^{\prime }\right\rangle }{1+a^{2}\tilde{\theta}^{2}l\left(
l+1\right) \mathfrak{g}_{l}^{2}\left( a,a\right) }\mathfrak{g}_{l}\left(
a,r^{\prime }\right) ,
\label{g0n-S3}
\end{equation}%
which we insert back into Eq. (\ref{g0n-S2}) with the final result  
\begin{equation}
g_{lmm^{\prime },\nu }^{0}\left( r,r^{\prime }\right) =\eta _{\phantom{0}\nu
}^{0}\delta _{mm^{\prime }}\left[ \mathfrak{g}_{l}\left( r,r^{\prime
}\right)-a^{2}\tilde{\theta}^{2}l\left( l+1\right) S_{l}\left( r,r^{\prime
}\right) \right] +ia\tilde{\theta}S_{l}\left( r,r^{\prime }\right)
\left\langle lm\right| \eta _{\phantom{k}\nu }^{k}\hat{\mathbf{L}}_{k}\left| lm^{\prime
}\right\rangle.
\label{g0n-S4}
\end{equation}%
where the function $S_{l}\left( r,r^{\prime }\right) $ was defined in Eq. (%
\ref{A-func}).

The remaining components now can be computed directly. The substitution of
Eq. (\ref{g0n-S3}) into Eq. (\ref{gkn-S}) produces 
\begin{eqnarray}
g_{lmm^{\prime },\nu }^{k}\left( r,r^{\prime }\right) &=& \eta _{\phantom{k}\nu }^{k}%
\mathfrak{g}_{l}\left( r,r^{\prime }\right) \delta _{mm^{\prime }}+ia\tilde{%
\theta}S_{l}\left( r,r^{\prime }\right) \eta _{\phantom{0}\nu }^{0}\left\langle
lm\right| \hat{\mathbf{L}}_{k}\left| lm^{\prime }\right\rangle -\nonumber \\
&& a^{2}\tilde{%
\theta}^{2}l\left( l+1\right) \mathfrak{g}_{l}\left( a,a\right) S_{l}\left(
r,r^{\prime }\right) \left\langle lm\right| \eta _{\phantom{r}\nu }^{r}\hat{\mathbf{L}%
}_{k}\hat{\mathbf{L}}_{r}\left| lm^{\prime }\right\rangle.
\label{gkn-S2}
\end{eqnarray}%
One can further check that $g_{lmm^{\prime },k}^{0}\left( r,r^{\prime
}\right) =g_{lmm^{\prime },0}^{k}\left( r,r^{\prime }\right) $. Thus the
general solution can be written in a compact way  as 
\begin{eqnarray}
g_{lmm^{\prime },\nu }^{\mu }\left( r,r^{\prime }\right) &= &\eta _{\phantom{\mu}\nu
}^{\mu }\mathfrak{g}_{l}\left( r,r^{\prime }\right) \delta _{mm^{\prime
}}-a^{2}\tilde{\theta}^{2}l\left( l+1\right) \mathfrak{g}_{l}\left(
a,a\right) S_{l}\left( r,r^{\prime }\right) \left\langle lm\right| \hat{%
\mathbf{L}}_{\mu }\hat{\mathbf{L}}_{\nu }\left| lm^{\prime }\right\rangle + \nonumber \\ 
&& ia%
\tilde{\theta}S_{l}\left( r,r^{\prime }\right) \left( \eta _{\phantom{\mu}0}^{\mu
}\Gamma _{\phantom{\alpha}\nu }^{\alpha }+\Gamma ^{\mu \alpha }\eta _{\phantom{0}\nu }^{0}\right)
\left\langle lm\right| \hat{\mathbf{L}}_{\alpha }\left| lm^{\prime
}\right\rangle
\label{c-RGF-S} 
\end{eqnarray}%
where $\hat{\mathbf{L}}_{0}$ denotes the identity operator and $\Gamma ^{\mu
\nu }=\eta ^{\mu \nu }-\eta _{\phantom{\mu}0}^{\mu }\eta _{\phantom{\nu}0}^{\nu }$.

\section{GF for a cylindrical $\protect\theta$ boundary}

\label{cyl_Green}

Now we concentrate in constructing the GF in cylindrical coordinates for the
configuration shown in Fig. \ref{cylfig} where the $\theta $ boundary is the
surface of a cylinder of radius $a$ with its axis lying along the $z$
direction . The various components of the GF are the solution of 
\begin{equation}
\left[ -\eta _{\phantom{\mu}\nu }^{\mu }\nabla ^{2}-\tilde{\theta}\delta \left( \rho
-a\right) n_{\alpha }\epsilon _{\phantom{\alpha \mu \beta} \nu }^{\alpha \mu \beta }\partial
_{\beta }\right] G_{\phantom{\nu}\sigma }^{\nu }\left( \mathbf{x},\mathbf{x}^{\prime
}\right) =4\pi \eta _{\phantom{\mu}\sigma }^{\mu }\delta \left( \mathbf{x}-\mathbf{x}%
^{\prime }\right) ,  \label{GE-C}
\end{equation}%
where $n_{\alpha }=\left( 0,\cos \varphi ,\sin \varphi ,0\right) $ is the
normal to the $\theta $ interface. Since we have  the completeness relation 
\begin{equation}
\delta \left( \varphi -\varphi ^{\prime }\right) \delta \left( z-z^{\prime
}\right) =\int_{-\infty }^{+\infty }\frac{dk}{2\pi }e^{ik\left( z-z^{\prime
}\right) }\frac{1}{2\pi }\sum_{m=-\infty }^{+\infty }\sum_{m^{\prime
}=-\infty }^{+\infty }\delta _{mm^{\prime }}e^{i\left( m\varphi -m^{\prime
}\varphi ^{\prime }\right) },  \label{C-Comp}
\end{equation}%
we look for a solution of the form 
\begin{equation}
G_{\phantom{\mu}\nu }^{\mu }\left( \mathbf{x},\mathbf{x}^{\prime }\right) =4\pi
\int_{-\infty }^{+\infty }\frac{dk}{2\pi }e^{ik\left( z-z^{\prime }\right) }%
\frac{1}{2\pi }\sum_{m=-\infty }^{+\infty }\sum_{m^{\prime }=-\infty
}^{+\infty }g_{mm^{\prime },\nu }^{\mu }\left( \rho ,\rho ^{\prime
};k\right) e^{i\left( m\varphi -m^{\prime }\varphi ^{\prime }\right) }.
\label{GF-C}
\end{equation}%
where $g_{mm^{\prime },\nu }^{\mu }\left( \rho ,\rho ^{\prime };k\right) $
is the reduced GF analogous to $g_{\phantom{\mu}\nu }^{\mu }\left( z,z^{\prime
}\right) $ in the case of planar symmetry.

In the limiting case $\tilde{\theta}\rightarrow 0$ the matrix elements take
the simple form $g_{mm^{\prime },\nu }^{\mu }\left( \rho ,\rho ^{\prime
};k\right) =\eta _{\phantom{\mu}\nu }^{\mu }\delta _{mm^{\prime }}\mathfrak{g}%
_{m}\left( \rho ,\rho ^{\prime };k\right) $, where $\mathfrak{g}_{m}\left(
\rho ,\rho ^{\prime };k\right) $ solves 
\begin{equation}
\hat{\mathcal{O}}_{\rho }^{(m)}\mathfrak{g}_{m}\left( \rho ,\rho ^{\prime
};k\right) =\frac{\delta \left( \rho -\rho ^{\prime }\right) }{\rho },
\label{O-Op-C}
\end{equation}%
with  the radial operator being 
\begin{equation}
\hat{\mathcal{O}}_{\rho }^{(m)}=-\frac{1}{\rho }\frac{\partial }{\partial
\rho }\left( \rho \frac{\partial }{\partial \rho }\right) +\frac{m^{2}}{\rho
^{2}}+k^{2}.  \label{O-Op-C2}
\end{equation}%
The solution to Eq. (\ref{O-Op-C}) in free space, with standard  boundary
conditions at infinity,  is 
\begin{equation}
\mathfrak{g}_{m}\left( \rho ,\rho ^{\prime };k\right) =\mbox{I}_{m}\left(
k\rho _{<}\right) \mbox{K}_{m}\left( k\rho _{>}\right) ,  \label{free-RGF-C}
\end{equation}%
where $\rho _{>}$ ($\rho _{<}$) is the greater (lesser) of $\rho $ and $\rho
^{\prime }$. Here $\mbox{I}_{m}$ and $\mbox{K}_{m}$ are the modified Bessel
functions of the first and second kind, respectively. The substitution of 
Eq. (\ref{free-RGF-C}) into Eq. (\ref{GF-C}) correctly reproduces the well-known
result $|\mathbf{x}-\mathbf{x}^{\prime }|^{-1}$ for the free case.

In the following we focus in solving Eq. (\ref{GE-C}) for the various
components of the GF matrix. To this end we first observe that the
additional differential operator in Eq. (\ref{GE-C}) does not involve radial
derivatives, but only derivatives with  respect to $\ $the coordinates $z$
and  $\varphi $. That is to say  
\begin{equation}
n_{\alpha }\epsilon _{\phantom{\alpha \mu \beta}\nu }^{\alpha \mu \beta }\partial _{\beta
}=\left( \cos\varphi \, \epsilon _{\phantom{1 \mu 3} \nu }^{1\mu 3}+\sin\varphi
\, \epsilon _{\phantom{2 \mu 3}\nu }^{2\mu 3}\right) \partial _{z}+\epsilon _{\phantom{1 \mu 2}
\nu }^{1\mu 2}\, \frac{1}{\rho }\partial _{\varphi }.  \label{O-Op-C3}
\end{equation}

Substituting Eq. (\ref{GF-C}) into Eq. (\ref{GE-C}) and using $\partial _{z}
\rightarrow i k$, $\partial _{\varphi} \rightarrow i m$, $- \nabla ^{2}
\rightarrow \hat{\mathcal{O}} ^{(m)} _{\rho}$ gives 
\begin{align}
\int _{- \infty} ^{+ \infty} \frac{dk}{2 \pi} e ^{ik \left( z - z ^{\prime}
\right)} \frac{1}{2 \pi} \sum _{m, m ^{\prime} = - \infty} ^{+ \infty} e ^{i
\left( m \varphi - m ^{\prime} \varphi ^{\prime} \right)} \left\lbrace \eta
_{\phantom{\mu}\nu }^{\mu} \hat{\mathcal{O}} _{\rho} ^{(m)} - i \tilde{\theta} \delta
\left( \rho - a \right) \left[ k \left( \cos\varphi\, \epsilon ^{1 \mu 3}
_{\phantom{1 \mu 3}\nu} + \sin\varphi\, \epsilon ^{2 \mu 3} _{\phantom{2 \mu 3}\nu} \right) +
\epsilon^{1 \mu 2} _{\phantom{1 \mu 2}\nu} \, \frac{m}{\rho} \right] \right\rbrace g
^{\nu} _{mm^{\prime},\sigma}  \notag \\
= \eta _{\phantom{\mu}\sigma}^{\mu } \frac{\delta \left( \rho - \rho ^{\prime} \right)%
}{\rho} \int _{- \infty} ^{+ \infty} \frac{dk}{2 \pi} e ^{ik \left( z - z
^{\prime} \right)} \frac{1}{2 \pi} \sum _{m , m ^{\prime} = - \infty} ^{+
\infty} \delta _{mm^{\prime}} e ^{i \left( m \varphi - m ^{\prime} \varphi
^{\prime} \right)}  \label{RGFE-C}
\end{align}

Using the linear independence of $e ^{- i m ^{\prime} \varphi ^{\prime} }$
and $e ^{- i k z ^{\prime} }$ we are left with 

\begin{eqnarray}
\sum _{m = - \infty}^{+ \infty} e^{i m \varphi } \left\lbrace \eta _{\phantom{\mu}
\nu}^{\mu} \hat{\mathcal{O}} _{\rho} ^{(m)} - i \tilde{\theta} \delta
\left( \rho - a \right) \left[ k \left( \cos\varphi \, \epsilon^{1 \mu 3}
_{\phantom{1 \mu 3}\nu} + \sin\varphi\, \epsilon^{2 \mu 3} _{\phantom{2 \mu 3}\nu} \right) +
\epsilon^{1 \mu 2} _{\phantom{1 \mu 2}\nu}\, \frac{m}{\rho} \right] \right\rbrace g
^{\nu} _{mm^{\prime},\sigma}&=&\eta _{\phantom{\mu}\sigma}^{\mu } \frac{\delta \left(
\rho - \rho ^{\prime} \right)}{\rho} \times \nonumber \\
&&\sum _{m = - \infty} ^{+ \infty} \delta
_{mm^{\prime}} e ^{i m \varphi}  \label{RGFE-C2}
\end{eqnarray}

In analogy with the spherical case, we next multiply Eq. (\ref{RGFE-C2}) to
the left by  $e^{-im^{\prime \prime }\varphi }$ and integrate with respect
to $\varphi $.  After using the the relations 
\begin{eqnarray}
\delta _{mm^{\prime }} &=&\frac{1}{2\pi }\int_{0}^{2\pi }e^{i\left(
m-m^{\prime }\right) \varphi }d\varphi ,  \label{ME-C} \\
A_{mm^{\prime \prime },\nu }^{\mu } &= &\frac{1}{2\pi }\int_{0}^{2\pi
}d\varphi e^{im\varphi }\left( \cos\varphi\, \epsilon _{\phantom{1 \mu 3} \nu }^{1\mu
3}+\sin\varphi \, \epsilon _{\phantom{2 \mu 3}\nu }^{2\mu 3}\right) e^{-im^{\prime
\prime }\varphi }=\frac{1}{2}\left[ \delta _{m,m^{\prime \prime }-1}
{({\tilde \epsilon}_{\phantom{\mu}\nu }^{\mu})}^\ast+\delta _{m,m^{\prime \prime }+1}\tilde{%
\epsilon}_{\phantom{\mu}\nu }^{\mu }\right] ,  \label{ME-C2}
\end{eqnarray}%
where $\tilde{\epsilon}_{\phantom{\mu}\nu }^{\mu }=\epsilon _{\phantom{1 \mu 3}\nu }^{1\mu
3}+i\epsilon_{\phantom{2 \mu 3}\nu }^{2\mu 3}$, Eq. (\ref{RGFE-C2}) simplifies
to 
\begin{equation}
\hat{\mathcal{O}}_{\rho }^{(m)}g_{mm^{\prime },\sigma }^{\mu }-i\tilde{\theta%
}\delta \left( \rho -a\right) \left[ k\sum_{m^{\prime \prime }=-\infty
}^{+\infty }A_{m^{\prime \prime }m,\nu }^{\mu }g_{m^{\prime \prime
}m^{\prime },\sigma }^{\nu }+\epsilon _{\phantom{1 \mu 2}\nu }^{1\mu 2}\,\frac{m}{\rho }%
\delta _{mm^{\prime }}g_{mm^{\prime },\sigma }^{\nu }\right] =\eta
_{\phantom{\mu}\sigma }^{\mu }\frac{\delta \left( \rho -\rho ^{\prime }\right) }{\rho }%
\delta _{mm^{\prime }},  \label{RGFE-C3}
\end{equation}%
where we have relabeled $m^{\prime \prime }\leftrightarrow m$.

The resulting equation can be integrated using the free
reduced GF $\mathfrak{g}_{m}\left( \rho ,\rho ^{\prime };k\right) $,
satisfying Eq. (\ref{O-Op-C}), with the result 
\begin{equation}
g_{mm^{\prime },\sigma }^{\mu }\left( \rho ,\rho ^{\prime }\right) =\eta
_{\phantom{\mu}\sigma }^{\mu }\delta _{mm^{\prime }}\mathfrak{g}_{m}\left( \rho ,\rho
^{\prime }\right) +i\tilde{\theta}m\epsilon _{\phantom{1 \mu 2}\nu }^{1\mu 2}\,
\mathfrak{g}_{m}\left( \rho ,a\right) g_{mm^{\prime },\sigma }^{\nu }\left(
a,\rho ^{\prime }\right) +i\tilde{\theta}ka\mathfrak{g}_{m}\left( \rho
,a\right) \sum_{m^{\prime \prime }=-\infty }^{+\infty }A_{m^{\prime \prime
}m,\nu }^{\mu }g_{m^{\prime \prime }m^{\prime },\sigma }^{\nu }\left( a,\rho
^{\prime }\right),
\label{RGFE-C4}
\end{equation}%
Note that we have suppressed the dependence of the reduced GF on $k$ for the
sake of brevity. Now we have to solve for the various components. To this
end we observe that the nonzero components of $A_{mm^{\prime \prime },\nu
}^{\mu }$ are 
\begin{eqnarray}
A_{mm^{\prime \prime },0}^{1} &=&A_{mm^{\prime \prime },1}^{0}=+\frac{i}{2}%
\left( \delta _{m,m^{\prime \prime }+1}-\delta _{m,m^{\prime \prime
}-1}\right) ,  \label{Amm-C} \\
A_{mm^{\prime \prime },0}^{2} &=&A_{mm^{\prime \prime },2}^{0}=-\frac{1}{2}%
\left( \delta _{m,m^{\prime \prime }+1}+\delta _{m,m^{\prime \prime
}-1}\right) .  \label{Amm-C2}
\end{eqnarray}%
This result allow us to split Eq. (\ref{RGFE-C4}) into the components $\mu =0
$, $\mu =3$ and $\mu =j=1,2$, obtaining  
\begin{align}
& g_{mm^{\prime },\sigma }^{0}\left( \rho ,\rho ^{\prime }\right) =\eta
_{\phantom{0}\sigma }^{0}\delta _{mm^{\prime }}\mathfrak{g}_{m}\left( \rho ,\rho
^{\prime }\right) +i\tilde{\theta}m\mathfrak{g}_{m}\left( \rho ,a\right)
g_{mm^{\prime },\sigma }^{3}\left( a,\rho ^{\prime }\right) +i\tilde{\theta}%
ka\mathfrak{g}_{m}\left( \rho ,a\right) \sum_{m^{\prime \prime }=-\infty
}^{+\infty }A_{m^{\prime \prime }m,i}^{0}g_{m^{\prime \prime }m^{\prime
},\sigma }^{i}\left( a,\rho ^{\prime }\right) ,  \label{g0n-C} \\
& g_{mm^{\prime },\sigma }^{3}\left( \rho ,\rho ^{\prime }\right) =\eta
_{\phantom{3}\sigma }^{3}\delta _{mm^{\prime }}\mathfrak{g}_{m}\left( \rho ,\rho
^{\prime }\right) +i\tilde{\theta}m\mathfrak{g}_{m}\left( \rho ,a\right)
g_{mm^{\prime },\sigma }^{0}\left( a,\rho ^{\prime }\right) ,  \label{g3n-C}
\\
& g_{mm^{\prime },\sigma }^{j}\left( \rho ,\rho ^{\prime }\right) =\eta
_{\phantom{j}\sigma }^{j}\delta _{mm^{\prime }}\mathfrak{g}_{m}\left( \rho ,\rho
^{\prime }\right) +i\tilde{\theta}ka\mathfrak{g}_{m}\left( \rho ,a\right)
\sum_{m^{\prime \prime }=-\infty }^{+\infty }A_{m^{\prime \prime
}m,0}^{j}g_{m^{\prime \prime }m^{\prime },\sigma }^{0}\left( a,\rho ^{\prime
}\right) .  \label{gin-C}
\end{align}%
where the index $i=1,2$.

Setting  $\rho =a$ in Eqs. (\ref{g3n-C}) and (\ref{gin-C}), and then 
substituting  into Eq. (\ref{g0n-C}) yields 
\begin{equation}
g_{mm^{\prime },\sigma }^{0}\left( \rho ,\rho ^{\prime }\right) =\eta
_{\phantom{0}\sigma }^{0}\delta _{mm^{\prime }}\mathfrak{g}_{m}\left( \rho ,\rho
^{\prime }\right) +i\tilde{\theta}\left[ m\delta _{mm^{\prime }}\eta
_{\phantom{3}\sigma }^{3}+kaA_{m^{\prime }m,\sigma }^{0}\right] \mathfrak{g}%
_{m}\left( \rho ,a\right) \mathfrak{g}_{m^{\prime }}\left( a,\rho ^{\prime
}\right) -\tilde{\theta}^{2}\mathfrak{g}_{m}\left( \rho ,a\right) \mathfrak{f%
}_{m}\left( k\right) g_{mm^{\prime },\sigma }^{0}\left( a,\rho ^{\prime
}\right),
\label{g0n-C2}
\end{equation}%
where $\mathfrak{f}_{m}\left( k\right) =m^{2}\mathfrak{g}_{m}\left(
a,a\right) +\frac{k^{2}a^{2}}{2}\left[ \mathfrak{g}_{m+1}\left( a,a\right) +%
\mathfrak{g}_{m-1}\left( a,a\right) \right] $. In deriving Eq. (\ref{g0n-C2}%
) we use the result 
\begin{equation}
\sum_{i=1}^{2}A_{mm^{\prime \prime },i}^{0}A_{m^{\prime }m,0}^{i}=\frac{1}{2}%
\left( \delta _{m,m^{\prime \prime }+1}\delta _{m^{\prime },m-1}+\delta
_{m,m^{\prime \prime }-1}\delta _{m^{\prime },m+1}\right) ,  \label{Amm-C3}
\end{equation}%
which can be verified directly from Eqs. (\ref{Amm-C}-\ref{Amm-C2}).

Solving for $g_{mm^{\prime },\sigma }^{0}\left( a,\rho ^{\prime }\right) $
by setting $\rho =a$ in Eq. (\ref{g0n-C2}) and inserting the result back in
this equation, we obtain 
\begin{equation}
g_{mm^{\prime },\sigma }^{0}\left( \rho ,\rho ^{\prime }\right) =\eta
_{\phantom{0}\sigma }^{0}\delta _{mm^{\prime }}\left[ \mathfrak{g}_{m}\left( \rho
,\rho ^{\prime }\right) -\tilde{\theta}^{2}\mathfrak{f}_{m}\left( k\right)
C_{mm}\left( \rho ,\rho ^{\prime }\right) \right] +i\tilde{\theta}\left(
m\delta _{mm^{\prime }}\eta _{\phantom{3}\sigma}^{3}+kaA_{m^{\prime }m,\sigma
}^{0}\right) C_{mm^{\prime }}\left( \rho ,\rho ^{\prime }\right) ,
\label{g0n-C3}
\end{equation}%
where the function $C_{mm^{\prime }}\left( \rho ,\rho ^{\prime }\right) $
was defined in Eq. (\ref{C-function}).

The remaining components now can be computed similarly. The substitution of $%
g_{mm^{\prime },\sigma }^{0}\left( a,\rho ^{\prime }\right) $ in Eqs. (\ref%
{g3n-C}) and (\ref{gin-C}) yields 
\begin{eqnarray}
g_{mm^{\prime },\sigma }^{3}\left( \rho ,\rho ^{\prime }\right)  &=&\eta
_{\phantom{3}\sigma }^{3}\delta _{mm^{\prime }}\left[ \mathfrak{g}_{m}\left( \rho
,\rho ^{\prime }\right) -m^{2}\tilde{\theta}^{2}\mathfrak{g}_{m}\left(
a,a\right) C_{mm}\left( \rho ,\rho ^{\prime }\right) \right] +im\tilde{\theta%
}\left( \eta_{\phantom{0}\sigma }^{0}+ika\tilde{\theta}A_{mm^{\prime },\sigma
}^{0}\right) C_{mm^{\prime }}\left( \rho ,\rho ^{\prime }\right) ,
\label{g3n-C2} \\
g_{mm^{\prime },\sigma }^{i}\left( \rho ,\rho ^{\prime }\right)  &=&\eta
_{\phantom{i}\sigma }^{i}\delta _{mm^{\prime }}\mathfrak{g}_{m}\left( \rho ,\rho
^{\prime }\right) +ika\tilde{\theta}\left[ \eta _{\phantom{0}\sigma }^{0}+i\eta
_{\phantom{3}\sigma }^{3}m^{\prime }\mathfrak{g}_{m^{\prime }}\left( a,a\right) %
\right] A_{m^{\prime }m,0}^{i}C_{m^{\prime }m}\left( \rho ^{\prime },\rho
\right)   \label{gin-C2} \nonumber \\
&&-\tilde{\theta}^{2}k^{2}a^{2}\mathfrak{g}_{m}\left( \rho ,a\right)
\sum_{m^{\prime \prime }=-\infty }^{+\infty }A_{m^{\prime
}m,0}^{j}A_{m^{\prime }m^{\prime \prime },\sigma }^{0}C_{m^{\prime \prime
}m^{\prime }}\left( a,\rho ^{\prime }\right) .
\end{eqnarray}

These results establish Eqs. (\ref{g0nCyl-3}-\ref{ginCyl-2}).

\end{document}